\documentclass[journal,10pt]{IEEEtran}

\usepackage{mathtools}	% loads amsmath
\usepackage{amssymb}
\usepackage{amsthm}

\usepackage[utf8]{inputenc}

\usepackage{ifthen}

\usepackage{microtype}

\usepackage{bm}

\usepackage[draft,nomargin,marginclue,inline]{fixme}
\makeatletter
\renewcommand*\FXLayoutMarginClue[3]{%
  \marginpar[%
  \raggedleft\@fxuseface{margin}\textcolor{red}{\ignorespaces $ \Rightarrow $}]{%
    \raggedright\@fxuseface{margin}\textcolor{red}{\ignorespaces $ \Leftarrow $}}}
\makeatother	% change color and displayed text in margin
\usepackage{pgfplotstable}	% loads pgfplots, tikz
\pgfplotsset{
	discard if/.style 2 args={
        x filter/.append code={
            \edef\tempa{\thisrow{#1}}
            \edef\tempb{#2}
            \ifx\tempa\tempb
                
            \fi
        }
    },
    discard if not/.style 2 args={
        x filter/.append code={
            \edef\tempa{\thisrow{#1}}
            \edef\tempb{#2}
            \ifx\tempa\tempb
            \else
                
            \fi
        }
    }
}

\usepackage{cite}
\usepackage{hyperref}
\hypersetup{
    colorlinks = true,
    linkbordercolor = {white},
    citecolor = {black},
    linkcolor = {black},
    urlcolor = {black}
}
\makeatletter
\renewcommand*{\eqref}[1]{%
  \hyperref[{#1}]{\textup{\tagform@{\ref*{#1}}}}%
}
\makeatother
\usepackage{tumcolor}

\usepackage[acronym,shortcuts]{glossaries}
\newacronym{cnn}{CNN}{convolutional neural network}
\newacronym{cs}{CS}{compressive sensing}
\newacronym{dft}{DFT}{discrete Fourier transform}
\newacronym{mmd}{MMD}{maximum mean discrepancy}
\newacronym{mse}{MSE}{mean squared error}
\newacronym{omp}{OMP}{orthogonal matching pursuit}
\newacronym{rip}{RIP}{restricted isometry property}
\newacronym{rkhs}{RKHS}{reproducing kernel Hilbert space}
\newacronym{snr}{SNR}{signal-to-noise ratio}
\newacronym{ula}{ULA}{uniform linear array}

\usepackage{cleveref}

\usepackage{enumitem}

\usepackage{algorithm}
\usepackage{algorithmic}

\DeclareMathOperator*{\argmax}{arg\,max}
\DeclareMathOperator*{\argmin}{arg\,min}

\DeclareMathOperator{\expec}{E}

\newcommand*{\mc}[1]{\mathcal{#1}}	% calligraphic font
\newcommand*{\mr}[1]{\mathrm{#1}}	% roman font

\newcommand*{\C}{\mathbb{C}}

\newcommand*{\R}{\mathbb{R}}

\newcommand{\herm}{{\operatorname{H}}}
\newcommand{\tp}{{\operatorname{T}}}

\newlength{\leftstackrelawd}
\newlength{\leftstackrelbwd}
\def\leftstackrel#1#2{\settowidth{\leftstackrelawd}%
	{${{}^{#1}}$}\settowidth{\leftstackrelbwd}{$#2$}%
	\addtolength{\leftstackrelawd}{-\leftstackrelbwd}%
	\leavevmode\ifthenelse{\lengthtest{\leftstackrelawd>0pt}}%
	{\kern-.5\leftstackrelawd}{}\mathrel{\mathop{#2}\limits^{#1}}}

\newcommand{\mbA}{\bm{A}}
\newcommand{\mbB}{\bm{B}}
\newcommand{\mbC}{\bm{C}}

\newcommand{\mbF}{\bm{F}}
\newcommand{\mbG}{\bm{G}}

\newcommand{\mbI}{\bm{I}}

\newcommand{\mbP}{\bm{P}}

\newcommand{\mbV}{\bm{V}}

\newcommand{\mbX}{\bm{X}}

\newcommand{\mba}{\bm{a}}

\newcommand{\mbg}{\bm{g}}
\newcommand{\mbh}{\bm{h}}

\newcommand{\mbn}{\bm{n}}

\newcommand{\mbs}{\bm{s}}

\newcommand{\mbu}{\bm{u}}
\newcommand{\mbv}{\bm{v}}

\newcommand{\mbx}{\bm{x}}
\newcommand{\mby}{\bm{y}}
\newcommand{\mbz}{\bm{z}}

\newcommand{\mbPhi}{{\bm{\Phi}}}

\newcommand{\mbPsi}{{\bm{\Psi}}}

\newcommand{\mbdelta}{{\bm{\delta}}}

\newcommand{\mbzero}{{\bm{0}}}

\pgfplotsset{compat=1.15}
\usetikzlibrary{shapes.geometric}

\Crefname{figure}{Fig.}{Figs.}

\DeclareMathOperator{\imag}{j}
\DeclareMathOperator{\mmd}{MMD}
\DeclareMathOperator{\stack}{stk}

\newcommand{\htilde}{\tilde{\mbh}}
\newcommand{\lbcs}{LBCS }
\newcommand{\Ttrain}{T_{\mr{tr}}}
\newcommand{\Tval}{T_{\mr{val}}}

\newcommand{\legendinit}{Alg. 2 with Alg. 4 init.}
\newcommand{\legendlbcs}{Alg. 4}
\newcommand{\legendlearned}{Alg. 2}
\newcommand{\legendomp}{random}

\newcommand{\plotheight}{0.6\columnwidth}
\newcommand{\plotwidth}{0.95\columnwidth}
\newcommand{\plotheightdictionary}{0.9\columnwidth}
\newcommand{\plotwidthdictionary}{0.5\columnwidth}
\newcommand{\plotheightsparse}{0.8\columnwidth}
\newcommand{\plotwidthsparse}{0.5\columnwidth}
\newcommand{\plotheightdelta}{0.7\columnwidth}
\newcommand{\plotwidthdelta}{0.95\columnwidth}

\pgfplotsset{tick label style={font=\small},label style={font=\small},legend style={font=\footnotesize}}

\definecolor{myblack}{RGB}{70,70,70}
\definecolor{myblue}{RGB}{65,105,225}
\definecolor{mygreen}{RGB}{0,139,139}
\definecolor{myorange}{RGB}{255,150,0}
\definecolor{myred}{RGB}{255,69,0}

\newcommand{\lineWidth}{1.0pt}
\tikzset{lbcs1/.style={mark options={solid},color=myblue,  line width=\lineWidth, mark=+, mark size=1.5pt, dotted}}
\tikzset{learned1/.style={mark options={solid},color=myorange, line width=\lineWidth, mark=x, mark size=1.5pt, dashed}}
\tikzset{learned3/.style={mark options={solid},color=myred, line width=\lineWidth, mark=x, mark size=1.5pt, solid}}
\tikzset{omp1/.style={mark options={solid},color=myblack, line width=\lineWidth, mark=*, mark size=0.5pt, solid}}

\tikzset{style1/.style={mark options={solid},color=myblack, line width=\lineWidth, mark=*, mark size=1.0pt, dashed}}
\tikzset{style2/.style={mark options={solid},color=myblue, line width=\lineWidth, mark=+, mark size=2.5pt, dashed}}
\tikzset{style3/.style={mark options={solid},color=myorange, line width=\lineWidth, mark=x, mark size=2.5pt, dashed}}
\tikzset{style4/.style={mark options={solid},color=myblack, line width=\lineWidth, mark=*, mark size=1.0pt, solid}}
\tikzset{style5/.style={mark options={solid},color=myblue, line width=\lineWidth, mark=+, mark size=2.5pt, solid}}
\tikzset{style6/.style={mark options={solid},color=myorange, line width=\lineWidth, mark=x, mark size=2.5pt, solid}}

\begin{document}

\title{Learning a Compressive Sensing Matrix with Structural Constraints via Maximum Mean Discrepancy Optimization}

\author{Michael~Koller and
        Wolfgang~Utschick% <-this % stops a space
\thanks{The authors are with Professur für Methoden der Signalverarbeitung, Technische Universität München,
80333 München, Germany, e-mail: michael.koller@tum.de, utschick@tum.de}}

\markboth{}
{Koller \MakeLowercase{\textit{et al.}}}

\maketitle

%%%%%%%%%%%%%%%%%%%%%%%%%%%%%%%%%%%%%%%%%%%%%%%%%%%%%%%%%%%%%%%%%%%%%%%%%%%%%%%
%%%%%%%%%%%%%%%%%%%%%%%%%%%%%%%%%%%%%%%%%%%%%%%%%%%%%%%%%%%%%%%%%%%%%%%%%%%%%%%
\begin{abstract}
    We introduce a learning-based algorithm to obtain a measurement matrix for compressive sensing related recovery problems.
    The focus lies on matrices with a constant modulus constraint
    which typically represent a network of analog phase shifters in hybrid precoding/combining architectures.
    We interpret a matrix with restricted isometry property as a mapping of points from a high- to a low-dimensional hypersphere.
    We argue that points on the low-dimensional hypersphere, namely, in the range of the matrix,
    should be uniformly distributed to increase robustness against measurement noise.
    This notion is formalized in an optimization problem which uses one of the maximum mean discrepancy metrics in the objective function.
    Recent success of such metrics in neural network related topics motivate a solution of the problem based on machine learning.
    Numerical experiments show better performance than random measurement matrices that are generally employed in compressive sensing contexts.
    Further, we adapt a method from the literature to the constant modulus constraint.
    This method can also compete with random matrices and it is shown to harmonize well with the proposed learning-based approach
    if it is used as an initialization.
    Lastly, we describe how other structural matrix constraints, e.g., a Toeplitz constraint, can be taken into account, too.
\end{abstract}


\IEEEpeerreviewmaketitle

%%%%%%%%%%%%%%%%%%%%%%%%%%%%%%%%%%%%%%%%%%%%%%%%%%%%%%%%%%%%%%%%%%%%%%%%%%%%%%%
%%%%%%%%%%%%%%%%%%%%%%%%%%%%%%%%%%%%%%%%%%%%%%%%%%%%%%%%%%%%%%%%%%%%%%%%%%%%%%%
\section{Introduction}

\IEEEPARstart{I}{n} the context of possible technologies for up-and-coming cellular systems,
millimeter wave systems with a high number of antennas have been
proposed~\cite{PiKh11,RaSuMaZhAzWaWoScSaGu1313,SuRaHeNiRa14,BaRaHe14,BoHeLoMaPo14}.
One way of controlling hardware costs and power consumption is to connect the antennas to a smaller number of receiver chains,
which is done via an analog mixing network~\cite{AyRaAbPiHe14,LiXuDo14,AlAyHe14,MeRuGoAlHe16,ArFoSiFrCa18}.
For data transmission, digital precoding is combined with configuring the analog mixing network (analog precoding).

In order to design both of these precoding stages (hybrid precoding),
state information of the high-dimensional channel is necessary.
However, the presence of an analog mixing network makes channel estimation challenging
because the observation vector has a smaller dimension (number of receiver chains) than the channel vector (number of antennas) which is to be estimated.
To facilitate channel estimation in this situation, channel models can play an important role.
In particular, sparsity-related structural knowledge about channels in millimeter wave systems has been exploited,
e.g.,~\cite{BaHaSaNo10,AlLeHe15,KiLo15}.

This naturally leads to the application of \ac{cs} theory in channel estimation contexts
where the analog mixing network then plays the role of the measurement or sensing matrix.
Among the best known conditions suitable measurement matrices should fulfill is the \ac{rip} (e.g., \cite{bookFoRa13}).
Since deterministic constructions of measurement matrices with \ac{rip} are not known,
the \ac{cs} theory uses random matrices which enjoy favorable properties with high probability \cite{bookFoRa13}.

One challenge with analog mixing networks is that they cannot realize any random matrix.
Rather, the hardware is constrained to provide matrix entries with constant modulus, i.e.,
only the phase of the entries can be adapted.
Additionally, it can be argued that such phase shifter matrices might only be able to provide a finite number of different phases (quantized phases, e.g., \cite{VeVe10}),
thus making the realization of random matrices that are dictated by the \ac{cs} theory more difficult.

In this paper, we propose a learning-based approach to find a deterministic measurement matrix for channel estimation.
The main idea is to interpret a matrix with \ac{rip} as a matrix which (approximately) maps points from a high-dimensional hypersphere to a low-dimensional hypersphere.
Moreover, we suggest to choose the matrix such that in addition the points in its range (on the low-dimensional hypersphere) are uniformly distributed
in order to combat the influence of measurement noise, which typically exists in communications systems.

We use one of the \ac{mmd} metrics~\cite{GrBoRaScSm12} as learning cost function to formalize the idea.
\ac{mmd}-based cost functions have had considerable success in machine learning, e.g., for generative networks~\cite{LiSwZe15,LiChChYiPo17,DzRoGh15}.
In numerical experiments, the matrix obtained from the proposed algorithm is shown to compete with or outperform the approach based on random matrices.
It may be argued that such a deterministic matrix might be easier to realize via an analog mixing network than random matrices.

Furthermore, we adapt an algorithm from~\cite{BaLiScGoBoCe16} to the phase shifter constant modulus constraint
and demonstrate its good performance in numerical experiments as well.
In addition, this algorithm is shown to harmonize well with the learning-based approach of the current paper
by providing an initialization of the learning algorithm.

The rest of this paper is organized as follows.
\Cref{sec:preliminaries} introduces the signal model as well as \ac{cs} background.
The main idea including a brief introduction to \ac{mmd} is outlined in \Cref{sec:main}.
An overview of prior work and the modification of~\cite{BaLiScGoBoCe16} is presented in \Cref{sec:related_work}.
Finally, numerical experiments are conducted in \Cref{sec:numerical_experiments} which is followed by a conclusion.

\emph{Notation:}
We write $ \exp(\mbX), \sin(\mbX), \cos(\mbX) $ for element-wise application of the exponential, sine, and cosine functions to a matrix $ \mbX $.
Transpose and conjugate transpose of vectors $ \mbx \in \C^n $ are denoted as $ \mbx^\tp $ and $ \mbx^\herm $, respectively.
We use $ \langle \cdot, \cdot \rangle $ for the Euclidean inner product and $ \| \cdot \| $ for its corresponding norm.
Lastly, the complex normal probability distribution with mean $ \mbzero_m \in \C^m $ and noise covariance $ \sigma^2 \mbI_m \in \C^{m\times m} $
is written as $ \mc{CN}(\mbzero_m, \sigma^2 \mbI_m) $.

%%%%%%%%%%%%%%%%%%%%%%%%%%%%%%%%%%%%%%%%%%%%%%%%%%%%%%%%%%%%%%%%%%%%%%%%%%%%%%%
%%%%%%%%%%%%%%%%%%%%%%%%%%%%%%%%%%%%%%%%%%%%%%%%%%%%%%%%%%%%%%%%%%%%%%%%%%%%%%%
\section{Preliminaries}\label{sec:preliminaries}

In what follows, the signal model is
\begin{equation}\label{eq:yAhn}
    \mby = \mbA \mbh + \mbn
\end{equation}
where $ \mbA \in \C^{m\times n} $ is called measurement matrix (with $ m < n $), $ \mbh \in \C^n $ is called channel,
$ \mby \in \C^m $ is a measurement or an observation, and $ \mbn \sim \mc{CN}(\mbzero_m, \sigma^2 \mbI_m) $ corresponds to measurement noise.
The matrix $ \mbA $ represents an analog mixing network with phase shifters
so that the matrix entries $ [\mbA]_{k,l} $ have the form
\begin{equation}\label{eq:elements_of_A}
    [\mbA]_{k,l} = \frac{1}{\sqrt{m}} \exp(\imag \phi_{k,l})
\end{equation}
with a phase $ \phi_{k,l} \in \R $.
All entries have a constant modulus constraint, $ |[\mbA]_{k,l}| = 1/\sqrt{m} $,
which is why we refer to such a matrix as constant modulus matrix.

This paper provides two methods to determine a constant modulus measurement matrix $ \mbA $
for recovering $ \mbh $ from~\eqref{eq:yAhn}.
The main contribution is a novel learning-based algorithm.
Additionally, we adapt a method from the literature to the constant modulus constraint.
This method computes a different matrix $ \mbA $ for every \ac{snr},
whereas the learning-based algorithm yields a single matrix which can be employed for all \acp{snr}.
We define the \ac{snr} as
\begin{equation}\label{eq:snr}
    \mr{SNR} = \frac{\| \mbA \mbh \|^2}{\expec[\|\mbn\|^2]} = \frac{\| \mbA \mbh \|^2}{m \sigma^2},
\end{equation}
where the noise variance $ \sigma^2 $ is adjusted for every pair $ (\mbA, \mbh) $ to achieve the desired \ac{snr}.

Since the goal is to estimate the $ n $-dimensional channel from an $ m $-dimensional observation with $ m < n $,
the inverse problem~\eqref{eq:yAhn} generally has no unique solution.
However, it is well known from \ac{cs} theory that a unique solution can exist if additional structural information about the channels is available.
One channel model that was studied in this context (see, e.g., \cite{AlLeHe15,KiLo15}) can be written as
\begin{equation}\label{eq:channel_model}
    \mbh = \sum_{k=1}^p s_k \mba(\theta_k)
\end{equation}
which corresponds to a sum of $ p $ paths which arrive from angles $ \theta_k \in [0, 2\pi] $ with path gains $ s_k \in \C $ at the base station.
For a uniform linear array at the base station, the steering vector $ \mba(\theta_k) \in \C^n $ has the form
\begin{equation}
    \mba(\theta_k) = \frac{1}{\sqrt{n}}
    \begin{bmatrix}
        \exp(\imag 0 \cdot \theta_k) & \dots & \exp(\imag (n-1) \cdot \theta_k)
    \end{bmatrix}^\tp.
\end{equation}
We briefly highlight relevant parts of \ac{cs} theory in the next section
where we then explain how \ac{cs} is used with this channel model.

%%%%%%%%%%%%%%%%%%%%%%%%%%%%%%%%%%%%%%%%%%%%%%%%%%%%%%%%%%%%%%%%%%%%%%%%%%%%%%%
\subsection{Compressive Sensing}\label{sec:cs}

\ac{cs} studies the inverse problem~\eqref{eq:yAhn} for general (no constant modulus constraint) matrices $ \mbB \in \C^{m\times n} $.
In the case where $ \mbh = \mbs \in \C^n $ is a $ p $-sparse vector ($ p $ nonzero entries),
it can be recovered if $ \mbB $ has a sufficient \ac{rip}~\cite{bookFoRa13}.
A matrix $ \mbB \in \C^{m\times n} $ is said to have the \ac{rip} if the restricted isometry constant $ \delta \geq 0 $ is small in
\begin{equation}\label{eq:rip}
    (1-\delta) \| \mbs \|^2 \leq \| \mbB \mbs \|^2 \leq (1+\delta) \| \mbs \|^2
\end{equation}
for all $ p $-sparse vectors $ \mbs \in \C^n $ \cite{bookFoRa13}.

More generally, it may be assumed that the signals of interest, $ \mbh $,
allow for a sparse representation $ \mbh = \mbPsi \mbs $ in some basis $ \mbPsi \in \C^{n\times n} $
such that the \ac{rip} condition is formulated where $ \mbs $ is replaced with $ \mbPsi \mbs $.
Even more generally, $ \mbh $ may lie in an (infinite) union of subspaces or some abstract subset $ \mc{H} \subset \C^n $
(see, e.g., \cite{ElMi09,Bl11,WiWeUt17})
and one defines a generalized notion of \ac{rip} as
\begin{equation}\label{eq:rip_general}
    (1-\delta) \| \mbh \|^2 \leq \| \mbB \mbh \|^2 \leq (1+\delta) \| \mbh \|^2
\end{equation}
for all $ \mbh \in \mc{H} $.

\begin{algorithm}[!t]
    \caption{Orthogonal Matching Pursuit (OMP)~\cite{bookFoRa13}}
    \label{alg:omp}
    \begin{algorithmic}[1]
        \REQUIRE matrix $ \mbC $, observation $ \mby $, sparsity $ p $
        \STATE $ S^{(0)} \leftarrow \emptyset $, $ \mbs^{(0)} \leftarrow \mbzero $
        \FOR {$ i = 1 $ to $ p $}
            \STATE $ j^* \leftarrow \argmax_{j\in\{1, \dots, n\}}\{[\mbC^\herm (\mby - \mbC \mbs^{(i-1)})]_j\} $
            \STATE $ S^{(i)} \leftarrow S^{(i-1)} \cup \{ j^* \} $
            \STATE $ \mbs^{(i)} \leftarrow \argmin_{\tilde{\mbs}}\{ \| \mby - \mbC \tilde{\mbs} \|,\, \mr{support}(\tilde{\mbs}) \subset S^{(i)} \} $
        \ENDFOR
        \RETURN $ \mbs^{(p)} \quad $ ($ p $-sparse vector)
    \end{algorithmic}
\end{algorithm}

One way of translating the model~\eqref{eq:channel_model} into a \ac{cs} setting is to introduce a so-called dictionary
\begin{equation}\label{eq:dictionary_grid}
    \mbPsi_L =
    \begin{bmatrix}
        \mba(\hat{\theta}_1) & \dots & \mba(\hat{\theta}_L)
    \end{bmatrix} \in \C^{n\times L}
\end{equation}
which collects $ L $ steering vectors corresponding to a grid of
$ L $ equidistantly sampled angles $ \hat{\theta}_\ell $ in $ [0, 2\pi] $.
Then, one approximates $ \mbh \approx \mbPsi_L \mbs $ with a $ p $-sparse vector $ \mbs \in \C^L $ which contains the path gains.
Channel estimation can now be performed by writing
\begin{equation}\label{eq:ytilde}
    \tilde{\mby} = (\mbA \mbPsi_L) \mbs + \mbn
\end{equation}
and by using for example \ac{omp} \cite{DaMaAv97,PaReKr93,Tr04} to recover an estimate $ \hat{\mbs} $ of $ \mbs $
so that the channel is estimated as $ \hat{\mbh} = \mbPsi_L \hat{\mbs} $.
\ac{omp} is described in \Cref{alg:omp} and would be used with $ \mbC = \mbA \mbPsi_L $, see, e.g., \cite{AlLeHe15}.
Note that~\eqref{eq:ytilde} is just an auxiliary approximation of~\eqref{eq:yAhn}
because $ \mbPsi_L \mbs $ is an approximation of $ \mbh $.
In particular, channels of the form~\eqref{eq:channel_model} do not have an exact representation
in terms of $ \mbPsi_L $ whenever $ \theta_k \notin \{ \hat{\theta}_1, \dots, \hat{\theta}_L \}$
for some $ k = 1, \dots, p $.

To obtain a matrix with \ac{rip}, one usually resorts to drawing random matrices which can fulfill the property with high probability~\cite{bookFoRa13}.
A typical example is a Gaussian random matrix $ \mbB \in \C^{m\times n} $
where the entries are drawn independently with $ [\mbB]_{k,l} \sim \mc{CN}(0, \frac{1}{\sqrt{m}}) $.
These, however, cannot be realized via an analog mixing network.
More generally, sub-Gaussian random matrices have been shown to fulfill the \ac{rip}~\cite{bookFoRa13}.
Sub-Gaussian random variables include bounded random variables~\cite{bookFoRa13}.
Thus, it makes sense to consider random constant modulus matrices $ \mbA $
which are randomly drawn with independent entries of the form
\begin{equation}\label{eq:random_Steinhaus}
    \frac{1}{\sqrt{m}} \exp(\imag \phi) \quad\text{with}\quad \phi \sim \mc{U}([0, 2\pi])
\end{equation}
where $ \phi $ is uniformly distributed in the interval $ [0, 2\pi] $.

%%%%%%%%%%%%%%%%%%%%%%%%%%%%%%%%%%%%%%%%%%%%%%%%%%%%%%%%%%%%%%%%%%%%%%%%%%%%%%%
%%%%%%%%%%%%%%%%%%%%%%%%%%%%%%%%%%%%%%%%%%%%%%%%%%%%%%%%%%%%%%%%%%%%%%%%%%%%%%%
\section{Learning a Measurement Matrix}\label{sec:main}

To motivate the main idea of this paper, we first rewrite the \ac{rip} condition~\eqref{eq:rip_general} in the form
\begin{equation}\label{eq:rip_divided}
    (1-\delta) \leq \left\| \mbB \frac{\mbh}{\|\mbh\|} \right\|^2 \leq (1+\delta)
\end{equation}
for all $ \mbh \in \mc{H} $.
Recovery algorithms need $ \delta \geq 0 $ to be small enough.
In the best case, we have $ \delta = 0 $ and~\eqref{eq:rip_divided} reads
\begin{equation}\label{eq:rip_hypersphere}
    \left\| \mbB \frac{\mbh}{\|\mbh\|} \right\|^2 = 1.
\end{equation}
Here, $ \htilde := \frac{\mbh}{\|\mbh\|} $ is a vector on the (high-dimensional) hypersphere in $ \C^n $
and according to~\eqref{eq:rip_hypersphere} the matrix $ \mbB $ has the property
that the vector $ \mbB \htilde $ lies on the (low-dimensional) hypersphere in $ \C^m $.
In the general case, in order for $ \mbdelta $ to be as small as possible,
\eqref{eq:rip_divided} requires the norm $ \| \mbB \htilde \| $ to be as close to one as possible.
In other words, we want to find a matrix which brings $ \mbB \htilde $ as close to the
hypersphere in $ \C^m $ as possible.

Let $ \mbh, \mbh' \in \mc{H} $ be two different vectors
and let $ \mby = \mbB \mbh $ and $ \mby' = \mbB \mbh' $ be the corresponding noiseless observations.
The~\ac{rip} formalizes the notion that the distance between $ \mby $ and $ \mby' $
should be proportional to the distance between $ \mbh $ and $ \mbh' $
such that two vectors $ \mbh, \mbh' $ which are far apart from each other
do not lead to the same observation even if the observations are noisy.
As a consequence, $ \mbh $ and $ \mbh' $ can still be uniquely recovered if the noise is small.
In this sense, the~\ac{rip} guarantees some robustness against measurement noise.

In the spirit of the above description,
it seems sensible in general to try to find a matrix $ \mbB $
which places two different observations far apart from each other,
so that even in the presence of noise, they are distinguishable.
When we work with the normalized vectors $ \htilde $,
this means that we now have two goals:
On the one hand, $ \| \mbB \htilde \| $ should be close to one in order to achieve a \ac{rip},
on the other hand, two different $ \mbB \htilde $ and $ \mbB \htilde' $ should be far apart
from each other.
The latter goal can also be formulated as follows:
We want to find a matrix $ \mbB $ which distributes $ \mbB \htilde $ isotropically around the origin.

To cast this idea into an optimization problem, let $ \mbu $ be a random variable which is uniformly distributed on the hypersphere in $ \C^m $.
That is, we have $ \| \mbu \| = 1 $ and $ \mbu $ is isotropically distributed around the origin.
Further, let all elements in $ \mc{H} $ be realizations of a random variable $ \mbh $
and consider again the normalized $ \htilde $.
Then, if $ d $ denotes a (for now generic) distance between the distributions of $ \mbB \htilde $ and $ \mbu $,
the goal is to choose $ \mbB $ such that this distance is minimized.
Together with the original goal to find a suitable constant modulus matrix $ \mbA $, we arrive at the abstract optimization problem:
\begin{equation}\label{eq:abstract_problem}
    \min_{\mbA\in\C^{m\times n}, |[\mbA]_{k,l}| = \frac{1}{\sqrt{m}}} d(\mbA \htilde, \mbu).
\end{equation}
We make this notion more precise in \Cref{sec:learning}.
To this end, \Cref{sec:mmd} first reviews the concept of \acp{mmd}
which can be used to define a distance between probability distributions.

The optimization problem~\eqref{eq:abstract_problem} encodes three goals.
In order for $ \mbA \htilde $ to be uniformly distributed on the hypersphere,
the norm $ \| \mbA \htilde \| $ needs to be as close to one as possible
since the norm of $ \mbu $ always equals one.
This corresponds to finding a matrix which has the~\ac{rip}
with $ \delta $ in~\eqref{eq:rip_divided} as small as possible.
At the same time, $ \mbA \htilde $ needs to be isotropically distributed around the origin
because this is a property of $ \mbu $.
This corresponds to placing noiseless observations far apart from each other.
Therefore, problem~\eqref{eq:abstract_problem} simultaneously aims for a \ac{rip} condition as well as for robustness against noise.
Finally, due to the constraints, we obtain a constant modulus matrix.

%%%%%%%%%%%%%%%%%%%%%%%%%%%%%%%%%%%%%%%%%%%%%%%%%%%%%%%%%%%%%%%%%%%%%%%%%%%%%%%
\subsection{Maximum Mean Discrepancy (MMD)}\label{sec:mmd}

To state problem~\eqref{eq:abstract_problem} more precisely, a distance between the probability distributions of $ \mbA \htilde $ and $ \mbu $ is necessary.
Many distances, for example the Kullback-Leibler divergence, can only be evaluated
if closed-form expressions of the distributions are available or if the distributions are estimated beforehand.
In the general case, we have access to samples of the channel $ \mbh $, which are, e.g., obtained from a simulation environment or a measurement campaign.
Moreover, the distribution of $ \mbh $, and thus in particular also of $ \mbA \htilde $, is not known.
It is therefore more convenient to work with a distance which can be evaluated based on samples from the involved distributions,
i.e., samples of $ \mbA \htilde $ and $ \mbu $.
In the following, we introduce one of the \ac{mmd} distances which has this property.

Let $ (\mc{X}, d) $ be a metric space and let $ \mbx $ and $ \mby $ be two random variables with respective probability measures $ p $ and $ q $.
\Ac{mmd} provides the possibility to define a metric on the space of probability measures on $ \mc{X} $.
For a given set $ \mc{F} $ of functions $ f: \mc{X} \to \R $,
\ac{mmd} is defined as~\cite[Definition~2]{GrBoRaScSm12}
\begin{equation}\label{eq:mmd}
    \mmd_{\mc{F}}(p, q) = \sup_{f\in\mc{F}}( \expec_{\mbx\sim p}[f(\mbx)] - \expec_{\mby\sim q}[f(\mby)] ).
\end{equation}
Depending on how $ \mc{F} $ is chosen, \ac{mmd} can be a metric
and then we have $ \mmd_{\mc{F}}(p, q) = 0 $ if and only if $ p = q $, and $ \mmd_{\mc{F}}(p, q) > 0 $ otherwise.
An overview of suitable function sets $ \mc{F} $ can, e.g., be found in~\cite{SrGrFuScLa10}.

Even if \ac{mmd} is a metric, the evaluation of~\eqref{eq:mmd}, i.e., solving the optimization problem, might still be difficult.
One particularly interesting set $ \mc{F} $ in this regard is the unit ball in a \ac{rkhs}.
A brief introduction into \acs{rkhs} can be found in~\cite{GrBoRaScSm12}.
Every \ac{rkhs} is a Hilbert space which is associated with a kernel $ k: \mc{X} \times \mc{X} \to \R $.
And, conversely, every kernel defines an associated \ac{rkhs}.
Intuitively speaking, a kernel $ k $ can represent all functions which are an element of the \ac{rkhs}.
For this reason, the optimization w.r.t. the whole set $ \mc{F} $ in~\eqref{eq:mmd} can be expressed by means of the kernel only
and the computation of the supremum is avoided.

Among the best known kernels is the Gaussian one:
\begin{equation}\label{eq:kernel_gauss}
    k_\sigma: \R^{2m} \times \R^{2m} \to \R, \: (\mbx, \mby) \mapsto k_\sigma(\mbx, \mby) = e^{-\frac{1}{2 \sigma^2} \| \mbx - \mby \|^2}
\end{equation}
with a parameter $ \sigma > 0 $.
Using the Gaussian kernel, \eqref{eq:mmd} can be expressed as~\cite[Lemma~6]{GrBoRaScSm12}
\begin{multline}\label{eq:mmd_kernel}
    \mmd_k^2(p, q) = \expec_{\mbx, \mbx'}[k(\mbx, \mbx')] \\- 2 \expec_{\mbx, \mby}[k(\mbx, \mby)] + \expec_{\mby, \mby'}[k(\mby, \mby')]
\end{multline}
and it is a metric.
Here, $ \mbx $ and $ \mbx' $ are independent random variables with distribution $ p $,
and $ \mby $ and $ \mby' $ are independent random variables with distribution $ q $.
The Gaussian kernel is a typical choice when $ \mmd_k $ is used, as, e.g., \cite{LiSwZe15,DzRoGh15,ReZhLiLu16} demonstrate.
One reason is that the associated \ac{rkhs} enjoys the property of being universal which makes~\eqref{eq:mmd_kernel} a metric~\cite[Theorem~5]{GrBoRaScSm12}.

If only samples $ \{ \mbx_i \}_{i=1}^{M} $ and $ \{ \mby_j \}_{j=1}^N $ of $ \mbx \sim p $ and $ \mby \sim q $, respectively, are given,
\eqref{eq:mmd_kernel} allows for a biased estimate~\cite[Equation~(5)]{GrBoRaScSm12}:
\begin{multline}\label{eq:mmd_empirical}
    \mmd_k^2(\{ \mbx_i \}_{i=1}^{M}, \{ \mby_j \}_{j=1}^N) = \sum_{i=1}^{M} \sum_{j=1}^{M} \frac{k(\mbx_i, \mbx_j)}{M^2}
    \\- 2 \sum_{i=1}^{M} \sum_{j=1}^N \frac{k(\mbx_i, \mby_j)}{MN} + \sum_{i=1}^N \sum_{j=1}^N \frac{k(\mby_i, \mby_j)}{N^2}.
\end{multline}
Throughout the paper, we employ this estimate~\eqref{eq:mmd_empirical} which converges in probability to~\eqref{eq:mmd_kernel}
at a rate of $ \mathcal{O}((M+N)^{-\frac{1}{2}}) $, see \cite[Theorem~7]{GrBoRaScSm12}.
A detailed analysis of the metric can be found in~\cite{GrBoRaScSm12}.

Since the kernel~\eqref{eq:kernel_gauss} is differentiable, so is $ \mmd_k $.
This, together with the fact that only samples from the involved distributions are required,
makes~\eqref{eq:mmd_empirical} interesting for learning applications.
For example, \cite{LiSwZe15} uses $ \mmd_k $ as neural network cost function in the context of data generation.
There, a set of true images is given and the goal is to find a function which takes random noise as argument
and yields artificial new images as function values.
The function is chosen to be a neural network and the goal is to minimize $ \mmd_k $
between its output and the true images,
so that the neural network is able to generate photo-realistic images
which resemble the true ones.
The minimization problem is solved via stochastic gradient descent with impressive results.

%%%%%%%%%%%%%%%%%%%%%%%%%%%%%%%%%%%%%%%%%%%%%%%%%%%%%%%%%%%%%%%%%%%%%%%%%%%%%%%
\subsection{Learning}\label{sec:learning}

With the help of \ac{mmd}, we express~\eqref{eq:abstract_problem} as a learning problem.
To this end, we want to replace the notion of a generic distance $ d $ in~\eqref{eq:abstract_problem} between $ \mbA \htilde $ and $ \mbu $ with $ \mmd_k $.

For this purpose, we represent $ \mbA \htilde $ and $ \mbu $ in terms of real quantities because it also simplifies the implementation.
Let
\begin{equation}
    \stack: \C^m \to \R^{2m},\:\mbz \mapsto \stack(\mbz) = [\Re(\mbz)^\tp, \Im(\mbz)^\tp]^\tp
\end{equation}
denote stacking real and imaginary parts of a complex vector $ \mbz $ into a real vector $ \stack(\mbz) $.
Further, let $ \{ \htilde_i \}_{i=1}^{\Ttrain} $ and $ \{ \mbu_i \}_{i=1}^{\Ttrain} $ be training samples.
Here, $ \{ \htilde_i \}_{i=1}^{\Ttrain} $ is a sample of normalized channel data points
and $ \{ \mbu_i \}_{i=1}^{\Ttrain} $ a set of points drawn from a uniform distribution on the hypersphere in $ \C^m $.
The channel data points may either stem from a measurement campaign or may be artificially generated according to a channel model.
A random vector $ \mbu $ with uniform distribution on the hypersphere in $ \C^m $ is obtained via normalization of an isotropic Gaussian random vector, e.g.,
$ \mbu = \frac{\mbv}{\|\mbv\|} $ with $ \mbv \sim \mc{C}\mc{N}(\mbzero_m, \mbI_m) $, which means that sampling is simple.
We then express~\eqref{eq:abstract_problem} as follows:
\begin{equation}\label{eq:complex_problem}
    \min_{\substack{\mbA\in\C^{m\times n}\\ |[\mbA]_{k,l}| = \frac{1}{\sqrt{m}}}}
    \mmd_k^2( \{ \stack(\mbA \htilde_i) \}_{i=1}^{\Ttrain}, \{ \stack(\mbu_i) \}_{i=1}^{\Ttrain}).
\end{equation}

Next, we address the constant modulus constraint $ |[\mbA]_{k,l}| = 1/\sqrt{m} $
and express the optimization problem in terms of real optimization variables instead of $ \mbA \in \C^{m\times n} $.
Note that since every entry of $ \mbA \in \C^{m\times n} $ has to have the form $ [\mbA]_{k,l} = \frac{1}{\sqrt{m}} \exp(\imag \phi_{k,l}) $ with a phase $ \phi_{k,l} \in \R $,
the constant modulus matrix $ \mbA $ is completely parameterized by $ mn $ real parameters.
We collect these parameters in a matrix $ \mbPhi \in \R^{m\times n} $
such that we can write $ \mbA = \frac{1}{\sqrt{m}}\exp(\imag \mbPhi) $ where the exponential function is meant to be applied element-wise
and where the constraint $ |[\mbA]_{k,l}| = \frac{1}{\sqrt{m}} $ is already incorporated.
If also $ \sin $ and $ \cos $ are applied element-wise,
we can write
\begin{equation}
    \mbA = \Re(\mbA) + \imag \Im(\mbA) = \frac{1}{\sqrt{m}}\cos(\mbPhi) + \imag \frac{1}{\sqrt{m}}\sin(\mbPhi)
\end{equation}
so that we get from $ \mbA \htilde = (\Re(\mbA) + \imag \Im(\mbA))(\Re(\htilde) + \imag \Im(\htilde)) $:
\begin{equation}\label{eq:stack_Ah}
    \stack(\mbA \htilde)
    =
    \begin{bmatrix}
        \Re(\mbA \htilde) \\
        \Im(\mbA \htilde)
    \end{bmatrix}
    = 
    \begin{bmatrix}
        \frac{\cos(\mbPhi)}{\sqrt{m}} & - \frac{\sin(\mbPhi)}{\sqrt{m}} \\
        \frac{\sin(\mbPhi)}{\sqrt{m}} & \frac{\cos(\mbPhi)}{\sqrt{m}}
    \end{bmatrix}
    \begin{bmatrix}
        \Re(\htilde) \\ \Im(\htilde)
    \end{bmatrix}.
\end{equation}
\Cref{eq:stack_Ah} reveals that $ \stack(\mbA \htilde) $ only depends on $ \mbA \in \C^{m\times n} $ via the real matrix $ \mbPhi \in \R^{m\times n} $.
Let us express this as $ \stack(\mbA \htilde) = \stack(\mbA(\mbPhi) \htilde) $ so that we can rewrite~\eqref{eq:complex_problem} as a real optimization problem:
\begin{equation}\label{eq:real_problem}
    \min_{\mbPhi\in\R^{m\times n}}
    \mmd_k^2( \{ \stack(\mbA(\mbPhi) \htilde_i) \}_{i=1}^{\Ttrain}, \{ \stack(\mbu_i) \}_{i=1}^{\Ttrain}).
\end{equation}

In the literature, such optimization problems are solved via stochastic gradient descent,
e.g., \cite{LiSwZe15,DzRoGh15,ReZhLiLu16,GaHu18}.
Since gradient descent can be employed and seems to yield good results,
\ac{mmd}-based problems are considered to be easy to optimize~\cite{GaHu18}.
Accordingly, we also use stochastic gradient descent to solve problem~\eqref{eq:real_problem}.
The details are summarized in \Cref{alg:learning}.
General convergence guarantees are part of ongoing research.
The standard procedure in machine learning is
to employ some form of grid or random search over various hyperparameters of the
stochastic gradient descent algorithm in combination with random initializations
of the optimization parameters.
A validation data set is then used to choose the best solution.
Possible hyperparameter values for \Cref{alg:learning} are described in \Cref{sec:implementation_details}.

In every iteration of~\Cref{alg:learning}, $ \mmd_k^2 $ and its gradient need to be calculated.
With backpropagation, evaluating the gradient has the same order of computation time
as the forward pass \cite{bookGrWa08}.
To compute $ \mmd_k^2 $ for a batch of $ T $ samples,
the kernel is evaluated $ \mathcal{O}(T^2) $ times, see~\eqref{eq:mmd_empirical},
with vectors of dimension $ 2m $ (stacked real and imaginary parts).
Since $ \stack(\mbA \htilde) $ can be computed in $ \mathcal{O}(mn) $ time,
the overall computation time is $ \mathcal{O}(mnT + 2mT^2) $.
As an alternative, \cite[Lemma~14]{GrBoRaScSm12} could be used,
where a linear (in $ T $) version of \ac{mmd} is introduced
such that the computation time would be $ \mathcal{O}(mnT) $.% $ \mathcal{O}((mn+2m)T) $.

\begin{algorithm}[!t]
    \caption{Learning $ \mbA = \mbA(\mbPhi) $}
    \label{alg:learning}
    \begin{algorithmic}[1]
        \REQUIRE training data $ \{ (\htilde_i, \mbu_i) \}_{i=1}^{\Ttrain} $
        \STATE randomly initialize $ \mbPhi \in [0, 2\pi]^{m\times n} $
        \WHILE {termination criterion not met}
            \STATE draw a batch of $ T $ samples uniformly from the training data: $ \{ (\htilde_t, \mbu_t ) \}_{t=1}^T \subset \{ (\htilde_i, \mbu_i) \}_{i=1}^{\Ttrain} $
            \STATE compute the stochastic gradient
                \begin{equation*}
                    \mbg_{\mbPhi} = \frac{\partial}{\partial \mbPhi} \mmd_k^2( \{ \stack(\mbA(\mbPhi) \htilde_t) \}_{t=1}^T, \{ \stack(\mbu_t) \}_{t=1}^T )
                \end{equation*}
            \STATE update $ \mbPhi $ with a gradient algorithm using $ \mbg_{\mbPhi} $
        \ENDWHILE
    \end{algorithmic}
\end{algorithm}

%%%%%%%%%%%%%%%%%%%%%%%%%%%%%%%%%%%%%%%%%%%%%%%%%%%%%%%%%%%%%%%%%%%%%%%%%%%%%%%
%%%%%%%%%%%%%%%%%%%%%%%%%%%%%%%%%%%%%%%%%%%%%%%%%%%%%%%%%%%%%%%%%%%%%%%%%%%%%%%
\section{Related Work}\label{sec:related_work}

This section reviews prior work related to learning a measurement matrix and describes a modification of one of the presented methods
which allows us to employ it in the context of the signal model described in \Cref{sec:preliminaries}.

%%%%%%%%%%%%%%%%%%%%%%%%%%%%%%%%%%%%%%%%%%%%%%%%%%%%%%%%%%%%%%%%%%%%%%%%%%%%%%%
\subsection{Prior Work}

The authors of \cite{WeChFr07} consider the real-valued signal model $ \mby = \mbB \mbh + \mbn $
where $ \mbB \in \R^{m\times n} $ is a general real matrix.
Their \textit{Uncertain Component Analysis} determines the matrix as
\begin{equation}
    \mbB = \argmax_{\tilde{\mbB} \in \R^{m\times n}, \tilde{\mbB} \tilde{\mbB}^\tp = \mbI} \prod_{i} \mr{Pr}(\mbh_i \mid \mby_i; \tilde{\mbB})
\end{equation}
with given data points $ \mbh_i $ and $ \mby_i =\tilde{\mbB} \mbh_i $ and
where $ \mr{Pr}(\mbh_i \mid \mby_i; \tilde{\mbB}) $ is the posterior probability of the data.
An algorithmic solution based on two fixed-point equations is proposed.

In \cite{El07}, the signal model is $ \mby = \mbB \mbPsi \mbs $
with $ \mbB \in \R^{m\times n} $ and $ \mbPsi \in \R^{n\times L} $ ($ n < L $).
The goal is to minimize the $ t $-averaged mutual coherence $ \mu_t(\mbB \mbPsi) $ with resepect to $ \mbB $.
To compute $ \mu_t(\mbB \mbPsi) $, the columns of $ \mbB \mbPsi $ are normalized and
then the Gram matrix $ \mbG = (\mbB \mbPsi)^\tp (\mbB \mbPsi) $ is computed.
The average of all entries $ |[\mbG]_{ij}| \geq t $ is defined to be $ \mu_t(\mbB \mbPsi) $.
The proposed iterative solution algorithm proceeds in two alternating stages.
Starting from some initial $ \mbB^{(1)} $,
in iteration $ i $ the Gram matrix $ \mbG^{(i)} = (\mbB^{(i)} \mbPsi)^\tp (\mbB^{(i)} \mbPsi) $ is computed
and its entries are modified such that $ \mu_t(\mbB^{(i)} \mbPsi) $ decreases,
yielding a new matrix $ \hat{\mbG}^{(i)} $.
In order to decrease $ \mu_t(\mbB^{(i)} \mbPsi) $, a \textit{down-scaling factor} $ \gamma \in (0, 1) $,
which is chosen before the algorithm's main iterations start,
is used to \textit{shrink} all entries of $ \mbG^{(i)} $ whose absolute values are larger than $ t $ by multiplying them by $ \gamma $.
In the second stage, a decomposition $ \hat{\mbG}^{(i)} \approx (\mbB^{(i+1)} \mbPsi)^\tp (\mbB^{(i+1)} \mbPsi) $ is found
to obtain a new matrix $ \mbB^{(i+1)} $.
These two stages are alternated for a number of iterations.

Similar to~\cite{El07}, the authors of \cite{DuSa09} work with the Gram matrix $ \mbG \in \R^{L\times L} $.
Their goal is to choose $ \mbB \in \R^{m\times n} $ such that $ \mbG $ is as close to an identity matrix as possible.
To this end, $ \mbB $ is initialized randomly and then a KSVD-like algorithm consisting of $ m $ (number of rows of $ \mbB $)
steps is computed.
Additionally, an extension is proposed where also the dictionary $ \mbPsi $ can be optimized.

The goal of~\cite{SaBaCe13} is to choose $ \mbB $ such that a bi-Lipschitz condition is established.
For given data points $ \{ \mbh_i \}_{i=1}^{T} $, this condition is expressed by replacing $ \mbh $ in~\eqref{eq:rip_general} with differences $ \mbh_i - \mbh_j $.
Therefore, the authors first define the secant set
$ \mc{S} = \{ \frac{\mbh_i -\mbh_j}{\| \mbh_i - \mbh_j \|}, 1 \leq i < j \leq T \} $
and then express the bi-Lipschitz criterion for this set as $ | \| \mbB \mbv \|^2 - 1 | = |\mbv^\tp \mbB^\tp \mbB \mbv - 1| \leq \delta $ with $ \mbv \in \mc{S} $.
Using the Gram matrix $ \mbG = \mbB^\tp \mbB $,
this motivates the following optimization problem to find a matrix $ \mbB \in \R^{m\times n} $:
\begin{multline}\label{eq:lipschitz_op}
    \min_{\mbG\in\R^{n\times n}} \sup_{\mbv\in\mc{S}} | \mbv^\tp \mbG \mbv - 1|
    \\\text{s.t.}\quad \mbG \succeq 0, \quad \mr{rank}(\mbG) = r, \quad \mr{trace}(\mbG) = b.
\end{multline}
The solution algorithm is based on a relaxation of~\eqref{eq:lipschitz_op}.

A related problem is studied in~\cite{HeSaYiBa15}:
\begin{equation}
    \min_{\mbG\in\R^{n\times n}, \mbG = \mbG^\tp, \mbG \succeq 0} \mr{trace}(\mbG)
    \quad\text{s.t.}\,
    \sup_{\mbv\in\mc{S}} | \mbv^\tp \mbG \mbv - 1 | \leq \delta.
\end{equation}
Once a feasible $ \mbG^* $ is determined, the matrix $ \mbB $ is found
via an eigendecomposition of $ \mbG^* $.

An approach motivated by neural network learning is presented in~\cite{WuDiSaYuHoStRoKu18}.
The signal model is $ \mby = \mbB \mbs $
with nonnegative sparse vectors $ \mbs \in \R_+^n $ and $ \mbB \in \R^{m\times n} $.
Motivated by an $ \ell_1 $-norm minimization approach,
the authors propose to utilize the following projected subgradient update rule to learn $ \mbB $:
\begin{equation}\label{eq:pgd}
    \mbs^{(i+1)} = \mbs^{(i)} - \frac{\beta}{i} (\mbI - \mbB^\tp \mbB) \mr{sign}(\mbs^{(i)})
\end{equation}
with the initialization $ \mbs^{(1)} = \mbB^\tp \mby $ and $ \beta \in \R $.
Every gradient iteration~\eqref{eq:pgd} is interpreted as a layer and $ I $ such gradient layers are concatenated.
This is called gradient unrolling.
The final output and estimate for $ \mbs $ is given by $ \hat{\mbs} = \mr{ReLU}(\mbs^{(I+1)}) $,
where $ \mr{ReLU} $ denotes the usual rectified linear unit
which performs the operation $ \max(x, 0) $ on every element $ x $ of its argument.
Then, $ \mbB \in \R^{m\times n} $ and $ \beta \in \R $ are interpreted as learnable parameters
and stochastic gradient descent is employed to learn $ \mbB $ and $ \beta $.
Specifically, given training data $ \{ \mbs_i \}_{i=1}^T $, a stochastic gradient descent method is used to solve
\begin{equation}
    \min_{\mbB \in \R^{m\times n}, \beta\in\R} \frac{1}{T} \sum_{i=1}^T \| \mbs_i - \hat{\mbs}_i \|^2.
\end{equation}

This gradient unrolling method was again employed in~\cite{WuLiCh19} for channel state information feedback.
To this end, a channel $ \mbh \in \C^n $ is compressed as $ \mby = \mbB [ \Re(\mbh)^\tp, \Im(\mbh)^\tp ]^\tp \in \R^m $
with $ \mbB \in \R^{m\times 2n} $ such that the base station
can recover $ \mbh $ from the feedback signal $ \mby $ via compressive sensing methods.
A modification of~\cite{WuDiSaYuHoStRoKu18} for channel estimation is described in~\cite{WuCh20},
where real and imaginary parts of channels $ \mbh \in \C^n $ are treated as independent data points
such that a real matrix $ \mbB \in \R^{m\times n} $ can be learned.

%%%%%%%%%%%%%%%%%%%%%%%%%%%%%%%%%%%%%%%%%%%%%%%%%%%%%%%%%%%%%%%%%%%%%%%%%%%%%%%
\subsection{Learning-Based Compressive Subsampling (LBCS)}

\begin{algorithm}[!t]
    \caption{LBCS~\cite{BaLiScGoBoCe16}}
    \label{alg:lbcs}
    \begin{algorithmic}[1]
        \REQUIRE training data $ \{ \mbh_i \}_{i=1}^{\Ttrain} $
        \REQUIRE matrix $ \mbV \in \C^{n\times n} $, number $ m $ of rows to select
        \STATE normalize the training data: $ \{ \htilde_i \}_{i=1}^{\Ttrain} \leftarrow \left\{ \frac{\mbh_i}{\| \mbh_i \|} \right\}_{i=1}^{\Ttrain} $
        \FOR {$ r = 1 $ to $ n $}
            \STATE $ \alpha_r \leftarrow \sum_{i=1}^{\Ttrain} |\langle \mbv_r, \htilde_i \rangle|^2 \quad $ (with $ \mbv_r $ row $ r $ of $ \mbV $)
        \ENDFOR
        \STATE $ \Omega \leftarrow $ indices $ r $ which correspond to $ m $ largest $ \alpha_r $
        \RETURN $ \mbP_\Omega \mbV \quad $ (learned measurement matrix)
    \end{algorithmic}
\end{algorithm}

One challenge with the methods presented so far is that general real-valued matrices $ \mbB \in \R^{m\times n} $ are found
whereas we are seeking a complex-valued matrix $ \mbA \in \C^{m\times n} $ where all entries $ [\mbA]_{k,l} $ have the same absolute value.
This constant modulus constraint is the main challenge.
In particular, methods which are based on a decomposition of the form $ \mbG = \mbB^\tp \mbB $ are difficult to apply
since such a decomposition need not exist when the entries of $ \mbB $ are required to have a constant modulus.
However, a modification of the approach published in~\cite{BaLiScGoBoCe16} allows us to incorporate the constraint.

The approach investigated in~\cite{BaLiScGoBoCe16} is based on the model
\begin{equation}\label{eq:lbcs_signal_model}
    \mby = \mbP_{\Omega} \mbV \mbh
\end{equation}
where $ \mbV \in \C^{n\times n} $ is a basis (a unitary matrix, $ \mbV^\herm \mbV = \mbI $)
and $ \mbP_\Omega \in \C^{m\times n} $ performs subsampling by selecting $ m = |\Omega| $ rows of $ \mbV $.
Once the row index set $ \Omega \subset \{1, 2, \dots, n\} $ is chosen, the measurement matrix is given by $ \mbB = \mbP_{\Omega} \mbV \in \C^{m\times n} $.
One strategy is to select the indices $ \Omega $ randomly.
However, the authors of~\cite{BaLiScGoBoCe16} argue that one can improve
upon this by considering a data-based selection procedure.

A criterion for selecting $ \Omega $ for which an exact solution is provided in~\cite{BaLiScGoBoCe16} can be described as follows.
Let $ \{ \mbh_i \}_{i=1}^T $ be given training data with norm one.
Then, select those $ m $ rows $ \mbv_r \in \C^n $ of $ \mbV $ which maximize
$ \frac{1}{T} \sum_{i=1}^T \sum_{r\in\Omega}|\langle \mbv_r, \mbh_i\rangle|^2 $.
This corresponds to maximizing the average captured energy $ \| \mbP_\Omega \mbV \mbh_i \|^2 $ of the observed data.
We call the procedure \lbcs after the paper's title and it is outlined in \Cref{alg:lbcs}.

%%%%%%%%%%%%%%%%%%%%%%%%%%%%%%%%%%%%%%%%%%%%%%%%%%%%%%%%%%%%%%%%%%%%%%%%%%%%%%%
\subsection{Adaptation of LBCS}

Since~\eqref{eq:lbcs_signal_model} considers the noiseless case and since it is assumed that a unitary matrix is already given,
we suggest the following adaptation of the algorithm to obtain a matrix $ \mbA $
which can be employed in the context of the current paper.

Because we need a constant modulus matrix,
the first modification is to divide the entries of $ \mbV $ by their respective absolute values.
\Cref{alg:lbcs} can then be used to select $ m $ rows of this constant modulus matrix.
Second, we do a Monte Carlo search over different initializations of the unitary matrix $ \mbV $.
To explain, we draw a unitary matrix $ \mbV $ uniformly at random,
divide its entries by their absolute values,
and run \Cref{alg:lbcs} to obtain a constant modulus matrix $ \mbA $.
An algorithm for generating random unitary matrices can, e.g., be found in~\cite{Me06}.
Then, validation data $ \{ \mbh_i \}_{i=1}^{\Tval} $ is used to evaluate the channel
estimation performance of the current $ \mbA $ in terms of \ac{mse}.
The best matrix $ \mbA_{\mr{MC}} $ of $ I $ such iterations is then the final result of the Monte Carlo search.
The whole procedure is outlined in \Cref{alg:mc_lbcs}.

%%%%%%%%%%%%%%%%%%%%%%%%%%%%%%%%%%%%%%%%%%%%%%%%%%%%%%%%%%%%%%%%%%%%%%%%%%%%%%%
\subsection{Comparison between MMD-Based Learning and LBCS}

Both~\Cref{alg:learning} and~\Cref{alg:mc_lbcs} are data-based approaches to determine a matrix $ \mbA $.
In both cases, an outer loop over different initializations is used.
\Cref{alg:learning} would be combined with multiple random searches,
and in case of~\Cref{alg:mc_lbcs},
we evaluate~\Cref{alg:lbcs} multiple times for different randomly initialized uniform matrices.
These are the training phases of the respective algorithms.
A main difference is that~\Cref{alg:mc_lbcs} uses a validation \ac{mse} criterion
to choose the best matrix, which is dependent on the \ac{snr}.
Thus, in the numerical experiments in \Cref{sec:numerical_experiments},
we run the algorithm for every \ac{snr}.
This is in contrast to~\Cref{alg:learning} where a single matrix is determined
which can be employed for all \acp{snr}.
Another difference is that \Cref{alg:mc_lbcs} selects rows of a random initialization $ \mbV $,
but doesn't modify its initial entries.
In comparison, \Cref{alg:learning} continually updates a random initialization.

\begin{algorithm}[!t]
    \caption{Monte Carlo LBCS}
    \label{alg:mc_lbcs}
    \begin{algorithmic}[1]
        \REQUIRE training and validation data $ \{ \mbh_j \}_{j=1}^{\Ttrain} $ and $ \{ \mbh_j \}_{j=1}^{\Tval} $
        \REQUIRE $ \mr{SNR} $, number $ m $ of rows to select, iterations $ I $
        \STATE $ \mbA_{\mr{MC}} \leftarrow \mbzero $, $ \mr{MSE}_{\mr{MC}} \leftarrow \infty $
        \FOR {$ i = 1 $ to $ I $}
            \STATE draw a random unitary matrix $ \mbV \in \C^{n\times n} $
            \STATE divide all entries of $ \mbV $ by their absolute values
            \STATE normalize the rows of $ \mbV $
            \STATE $ \mbA^{(i)} \leftarrow \mr{LBCS}(\{ \mbh_j \}_{j=1}^{\Ttrain}, \mbV, m) $
            \STATE $ \mr{MSE}^{(i)} \leftarrow \mr{Evaluation}(\mbA^{(i)}, \{\mbh_j\}_{j=1}^{\Tval}, \mr{SNR}) $
            \IF {$ \mr{MSE}^{(i)} < \mr{MSE}_{\mr{MC}} $}
            \STATE $ \mr{MSE}_{\mr{MC}} \leftarrow \mr{MSE}^{(i)} $
                \STATE $ \mbA_{\mr{MC}} \leftarrow \mbA^{(i)} $
            \ENDIF
        \ENDFOR
        \RETURN $ \mbA_{\mr{MC}} $
    \end{algorithmic}
\end{algorithm}

%%%%%%%%%%%%%%%%%%%%%%%%%%%%%%%%%%%%%%%%%%%%%%%%%%%%%%%%%%%%%%%%%%%%%%%%%%%%%%%
%%%%%%%%%%%%%%%%%%%%%%%%%%%%%%%%%%%%%%%%%%%%%%%%%%%%%%%%%%%%%%%%%%%%%%%%%%%%%%%
\section{Numerical Experiments}\label{sec:numerical_experiments}

In numerical experiments, we cover the cases
(i) training with sparse vectors,
(ii) training with vectors that have an exact sparse representation,
and (iii) training with vectors that are only approximately sparse.
Channel estimation is always performed with \ac{omp} (\Cref{alg:omp}).
While there are other choices for recovery algorithms, like $ \ell_1 $-norm minimization~\cite{bookFoRa13},
the exact choice is not too critical
since we are only interested in the relative performance of the different approaches to determine a measurement matrix.
\ac{omp} is a cheap option and widely used.

%%%%%%%%%%%%%%%%%%%%%%%%%%%%%%%%%%%%%%%%%%%%%%%%%%%%%%%%%%%%%%%%%%%%%%%%%%%%%%%
\subsection{Data Models}
Of the following three data models, only the last would actually be considered a model for communications channels.
Nonetheless, for simplicity, we continue to speak of channels and channel estimation.
We investigate these three models because they represent typical cases of \ac{cs} applications.

First, channels
\begin{equation}\label{eq:model1}
    \mbh = \mbs
\end{equation}
are modeled as sparse vectors $ \mbs \in \C^{n} $ with $ p $ nonzero entries.
Second, channels are modeled as sparse with respect to the \ac{dft} basis $ \mbF \in \C^{n\times n} $:
\begin{equation}\label{eq:model2}
    \mbh = \mbF \mbs.
\end{equation}
In both cases, each of the $ p $ nonzero entries of $ \mbs $ is drawn from a Gaussian distribution $ \mc{CN}(0, \frac{1}{p}) $.
Third, channels are modeled as explained around~\eqref{eq:channel_model} such that we have
\begin{equation}\label{eq:model3}
    \mbh = \sum_{k=1}^p s_k \mba(\theta_k).
\end{equation}
The path gains $ s_k $ are drawn from $ \mc{CN}(0, \frac{1}{p}) $, the angles $ \theta_k $ are drawn uniformly in $ [0, 2\pi] $.
To perform channel estimation, we use the dictionary $ \mbPsi_L $ from~\eqref{eq:dictionary_grid} with $ L = 16 n $.

To estimate a channel $ \mbh $ from an observation $ \mby = \mbA \mbh + \mbn $,
\ac{omp} (\Cref{alg:omp}) is used with $ \mbC = \mbA \mbPsi $
where $ \mbPsi $ is chosen as (i) $ \mbPsi = \mbI $, (ii) $ \mbPsi = \mbF $, or (iii) $ \mbPsi = \mbPsi_L $,
depending on the channel model.
In all cases, \ac{omp} yields a $ p $-sparse vector $ \hat{\mbs} $ and the channel estimate is computed as $ \hat{\mbh} = \mbPsi \hat{\mbs} $.
In the following numerical experiments,
the described procedure is repeated for $ T_{\mr{test}} = 10000 $ randomly drawn channels
such that an overall relative \ac{mse} $ \frac{\expec[\|\hat{\mbh} - \mbh\|^2]}{\expec[\|\mbh\|^2]} $ can be estimated.
For a given measurement matrix $ \mbA $ determined via learning (\Cref{alg:learning}) or via \lbcs (\Cref{alg:mc_lbcs}),
the evaluation process is outlined in \Cref{alg:evaluation}.

\begin{algorithm}[!t]
    \caption{Evaluation}
    \label{alg:evaluation}
    \begin{algorithmic}[1]
        \REQUIRE measurement matrix $ \mbA $, data $ \{ \mbh_i \}_{i=1}^{T} $
        \REQUIRE $ \mr{SNR} $, sparsity $ p $, dictionary $ \mbPsi $
        \STATE $ e \leftarrow 0 $, $ \Delta \leftarrow 0 $
        \FOR {$ i = 1 $ to $ T $}
            \STATE get $ \sigma $ with $ (\mbA, \mbh_i) $ via~\eqref{eq:snr}, draw $ \mbn \sim \mc{CN}(\mbzero_m, \sigma^2 \mbI_m) $
            \STATE $ \mby \leftarrow \mbA \mbh_i + \mbn \quad $ (compute observation)
            \STATE $ \hat{\mbh} \leftarrow \mbPsi \mr{OMP}(\mbA\mbPsi, \mby, p) \quad $ (estimate channel $ \mbh_i $)
            \STATE $ e \leftarrow e + \| \mbh_i \|^2 $, $ \Delta \leftarrow \Delta + \| \hat{\mbh} - \mbh_i \|^2 \quad $ (squared error)
        \ENDFOR
        \RETURN $ \frac{\Delta}{e} \quad $ (relative mean squared error)
    \end{algorithmic}
\end{algorithm}

As already pointed out in \Cref{sec:related_work},
the related methods which are based on decompositions of a Gram matrix are difficult to employ
when the measurement matrix is constrained to have constant modulus entries
because a corresponding decomposition does generally not exist.
However, the method from~\cite{BaLiScGoBoCe16} could be adapted to the constraint
and is therefore used as a reference, see~\Cref{alg:mc_lbcs}.

We compare the two methods of determining a measurement matrix $ \mbA $ (\Cref{alg:learning,alg:mc_lbcs})
with the alternative approach of drawing a new random measurement matrix for every channel.
That is, instead of first fixing $ \mbA $ and then running \Cref{alg:evaluation},
the first step in the for-loop (between lines 2 and 3) always consists of drawing a new random measurement matrix
(with entries as explained in~\eqref{eq:random_Steinhaus})
such that as many measurement matrices are drawn as there are channels.
This approach is labeled as ``random'' in the following plots.

%%%%%%%%%%%%%%%%%%%%%%%%%%%%%%%%%%%%%%%%%%%%%%%%%%%%%%%%%%%%%%%%%%%%%%%%%%%%%%%
\subsection{Implementation Details}\label{sec:implementation_details}

Whenever we run \Cref{alg:mc_lbcs},
we use $ \Ttrain = 100000 $ training and $ \Tval = 1000 $ validation data samples which are randomly generated according to one of the three channel models.
Further, for every given \ac{snr}, we do a Monte Carlo search over $ I = 100 $ matrices.

In order to run the learning \Cref{alg:learning}, a number of hyperparameters need to be set.
First of all, since we use \ac{mmd} with a Gaussian kernel~\eqref{eq:kernel_gauss}, the parameter $ \sigma > 0 $ needs to be chosen.
As argued in~\cite{LiSwZe15}, finding the optimal $ \sigma $ which makes \ac{mmd} most efficient is an open problem,
but a mixture of kernels may serve the intended purpose.
Therefore, we use $ \mmd_k $ in~\eqref{eq:mmd_empirical} with the kernel
\begin{equation}
    k(x, y) = \sum_{\sigma\in\mc{S}} k_\sigma(x, y)
\end{equation}
where $ \mc{S} = \{2, 5, 10, 20, 40, 80\} $.

We use Pytorch \cite{pytorch} to implement the stochastic gradient optimization in \Cref{alg:learning} with the Adam optimizer \cite{KiBa14}.
There are three hyperparameters: batch size $ T $, learning rate (gradient step size) $ l_r $, and exponential learning rate decay $ \beta $.
As described in~\cite{BeBe12}, we determine these three hyperparameters via random search.
In detail, we draw $ (T, l_r, \beta) \in [150, 1500] \times [10^{-6}, 5\cdot 10^{-3}] \times [0.94, 1] $ randomly and run \Cref{alg:learning}.
This is repeated 64 times and validation data is then used to pick the best of the 64 so-obtained matrices $ \mbA $.
The number of training samples is always $ \Ttrain = 50000 $ in all cases and settings.
For reasons explained later in \Cref{sec:data_normalization}, training data points $ \{ \mbh_i \}_{i=1}^{\Ttrain} $ are normalized as follows:
\begin{equation}\label{eq:normalization_expec}
    \htilde_i = \frac{\mbh_i}{\sqrt{\frac{1}{\Ttrain}\sum_{j=1}^{\Ttrain} \| \mbh_j \|^2}}, \quad \forall i = 1, 2, \dots, \Ttrain.
\end{equation}
The termination criterion for \Cref{alg:learning} is early stopping~\cite{bookGoBeCo16}.

%%%%%%%%%%%%%%%%%%%%%%%%%%%%%%%%%%%%%%%%%%%%%%%%%%%%%%%%%%%%%%%%%%%%%%%%%%%%%%%
\subsection{Discussion of Results for Sparse Vectors}

We first concentrate on the two models~\eqref{eq:model1} and~\eqref{eq:model2} where channels have an exact sparse representation
(either in the canonical basis $ \mbI $ or in the \ac{dft} basis $ \mbF $).
The constant modulus measurement matrix $ \mbA \in \C^{m\times 128} $ has the dimensions $ m \times 128 $ and we vary the sparsity $ p \in \{1, 5, 10\} $.

%
% true_sparse and basis_sparse, m = 16
%
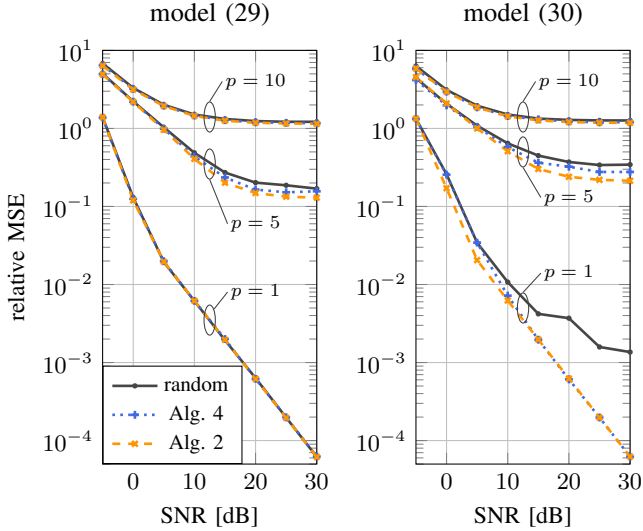
\begin{figure}[!t]
\centering
    \begin{tikzpicture}
        \begin{axis}[
            ymode=log,
            height=\plotheightsparse,
            width=\plotwidthsparse,
            ticks=both,
            xmin=-5, xmax=30,
            ymin=0.00005, ymax=10,
            xlabel={SNR [dB]},
            ylabel={relative MSE},
            title={model~\eqref{eq:model1}},
            xmajorgrids,
            ymajorgrids,
            legend style={at={(0.0, 0.0)}, anchor=south west, legend columns=1},
        ]

        % random
        % p = 1
        \addplot[omp1, discard if not={algorithm}{random}]
            table [ignore chars=",x=snr,y=mean,col sep=comma]
            {csv/compare_all/true_sparse/nAntennas_128_nPaths_1_nObservations_16.csv};
        \label{curve:random_p1_m16_true_sparse}
        \addlegendentry{\legendomp}

        % random
        % p = 5
        \addplot[forget plot, omp1, discard if not={algorithm}{random}]
            table [ignore chars=",x=snr,y=mean,col sep=comma]
            {csv/compare_all/true_sparse/nAntennas_128_nPaths_5_nObservations_16.csv};

        % random
        % p = 10
        \addplot[forget plot, omp1, discard if not={algorithm}{random}]
            table [ignore chars=",x=snr,y=mean,col sep=comma]
            {csv/compare_all/true_sparse/nAntennas_128_nPaths_10_nObservations_16.csv};

        % lbcs
        % p = 1
        \addplot[lbcs1, discard if not={algorithm}{lbcs}]
            table [ignore chars=",x=snr,y=mean,col sep=comma]
            {csv/compare_all/true_sparse/nAntennas_128_nPaths_1_nObservations_16.csv};
        \label{curve:lbcs_p1_m16_true_sparse}
        \addlegendentry{\legendlbcs}

        % lbcs
        % p = 5
        \addplot[forget plot, lbcs1, discard if not={algorithm}{lbcs}]
            table [ignore chars=",x=snr,y=mean,col sep=comma]
            {csv/compare_all/true_sparse/nAntennas_128_nPaths_5_nObservations_16.csv};

        % lbcs
        % p = 10
        \addplot[forget plot, lbcs1, discard if not={algorithm}{lbcs}]
            table [ignore chars=",x=snr,y=mean,col sep=comma]
            {csv/compare_all/true_sparse/nAntennas_128_nPaths_10_nObservations_16.csv};

        % learned
        % p = 1
        \addplot[learned1, discard if not={algorithm}{learned}]
            table [ignore chars=",x=snr,y=mean,col sep=comma]
            {csv/compare_all/true_sparse/nAntennas_128_nPaths_1_nObservations_16.csv};
        \addlegendentry{\legendlearned}

        % learned
        % p = 5
        \addplot[forget plot, learned1, discard if not={algorithm}{learned}]
            table [ignore chars=",x=snr,y=mean,col sep=comma]
            {csv/compare_all/true_sparse/nAntennas_128_nPaths_5_nObservations_16.csv};
        \label{curve:learned_p1_m16_true_sparse}

        % learned
        % p = 10
        \addplot[forget plot, learned1, discard if not={algorithm}{learned}]
            table [ignore chars=",x=snr,y=mean,col sep=comma]
            {csv/compare_all/true_sparse/nAntennas_128_nPaths_10_nObservations_16.csv};

        \draw (axis cs:12.5,1.4) node[ellipse, darkgray, inner xsep=0pt, inner ysep=0pt, minimum width=0.15cm, minimum height=0.4cm, draw] (e10) {};
        \node[inner sep=1pt] (n10) at (axis cs:20,4) {\scriptsize $ p = 10 $};
        \draw[darkgray] (n10.west) -- (e10.north);

        \draw (axis cs:12.5,0.35) node[ellipse, darkgray, inner xsep=0pt, inner ysep=0pt, minimum width=0.15cm, minimum height=0.4cm, draw] (e5) {};
        \node[inner sep=1pt] (n5) at (axis cs:20,0.06) {\scriptsize $ p = 5 $};
        \draw[darkgray] (n5.west) -- (e5.south);

        \draw (axis cs:12.5,0.0035) node[ellipse, darkgray, inner xsep=0pt, inner ysep=0pt, minimum width=0.15cm, minimum height=0.4cm, draw] (e1) {};
        \node[inner sep=1pt] (n1) at (axis cs:20,0.008) {\scriptsize $ p = 1 $};
        \draw[darkgray] (n1.west) -- (e1.north east);

        \end{axis}
    \end{tikzpicture}
    %%%%%%%%%%
    \begin{tikzpicture}
        \begin{axis}[
            ymode=log,
            height=\plotheightsparse,
            width=\plotwidthsparse,
            ticks=both,
            xmin=-5, xmax=30,
            ymin=0.00005, ymax=10,
            xlabel={SNR [dB]},
            title={model~\eqref{eq:model2}},
            xmajorgrids,
            ymajorgrids,
            legend style={at={(0.0, 0.0)}, anchor=south west, legend columns=1},
        ]

        % random
        % p = 1
        \addplot[omp1, discard if not={algorithm}{random}]
            table [ignore chars=",x=snr,y=mean,col sep=comma]
            {csv/compare_all/basis_sparse/nAntennas_128_nPaths_1_nObservations_16.csv};

        % random
        % p = 5
        \addplot[forget plot, omp1, discard if not={algorithm}{random}]
            table [ignore chars=",x=snr,y=mean,col sep=comma]
            {csv/compare_all/basis_sparse/nAntennas_128_nPaths_5_nObservations_16.csv};

        % random
        % p = 10
        \addplot[forget plot, omp1, discard if not={algorithm}{random}]
            table [ignore chars=",x=snr,y=mean,col sep=comma]
            {csv/compare_all/basis_sparse/nAntennas_128_nPaths_10_nObservations_16.csv};

        % lbcs
        % p = 1
        \addplot[lbcs1, discard if not={algorithm}{lbcs}]
            table [ignore chars=",x=snr,y=mean,col sep=comma]
            {csv/compare_all/basis_sparse/nAntennas_128_nPaths_1_nObservations_16.csv};

        % lbcs
        % p = 5
        \addplot[forget plot, lbcs1, discard if not={algorithm}{lbcs}]
            table [ignore chars=",x=snr,y=mean,col sep=comma]
            {csv/compare_all/basis_sparse/nAntennas_128_nPaths_5_nObservations_16.csv};

        % lbcs
        % p = 10
        \addplot[forget plot, lbcs1, discard if not={algorithm}{lbcs}]
            table [ignore chars=",x=snr,y=mean,col sep=comma]
            {csv/compare_all/basis_sparse/nAntennas_128_nPaths_10_nObservations_16.csv};

        % learned
        % p = 1
        \addplot[learned1, discard if not={algorithm}{learned}]
            table [ignore chars=",x=snr,y=mean,col sep=comma]
            {csv/compare_all/basis_sparse/nAntennas_128_nPaths_1_nObservations_16.csv};

        % learned
        % p = 5
        \addplot[forget plot, learned1, discard if not={algorithm}{learned}]
            table [ignore chars=",x=snr,y=mean,col sep=comma]
            {csv/compare_all/basis_sparse/nAntennas_128_nPaths_5_nObservations_16.csv};

        % learned
        % p = 10
        \addplot[forget plot, learned1, discard if not={algorithm}{learned}]
            table [ignore chars=",x=snr,y=mean,col sep=comma]
            {csv/compare_all/basis_sparse/nAntennas_128_nPaths_10_nObservations_16.csv};

        \draw (axis cs:12.5,1.4) node[ellipse, darkgray, inner xsep=0pt, inner ysep=0pt, minimum width=0.15cm, minimum height=0.4cm, draw] (e10) {};
        \node[inner sep=1pt] (n10) at (axis cs:20,4) {\scriptsize $ p = 10 $};
        \draw[darkgray] (n10.west) -- (e10.north);

        \draw (axis cs:12.5,0.5) node[ellipse, darkgray, inner xsep=0pt, inner ysep=0pt, minimum width=0.15cm, minimum height=0.4cm, draw] (e5) {};
        \node[inner sep=1pt] (n5) at (axis cs:20,0.13) {\scriptsize $ p = 5 $};
        \draw[darkgray] (n5.west) -- (e5.south);

        \draw (axis cs:12.5,0.005) node[ellipse, darkgray, inner xsep=0pt, inner ysep=0pt, minimum width=0.15cm, minimum height=0.4cm, draw] (e1) {};
        \node[inner sep=1pt] (n1) at (axis cs:20,0.014) {\scriptsize $ p = 1 $};
        \draw[darkgray] (n1.west) -- (e1.north);

        \end{axis}
    \end{tikzpicture}
\caption{%
    Evaluation with channel models~\eqref{eq:model1} (left) and~\eqref{eq:model2} (right).
    The constant modulus matrices $ \mbA $ have size $ 16 \times 128 $ and the sparsity $ p \in \{1, 5, 10\} $ is varied.
    Evaluation is shown with random matrices (\ref{curve:random_p1_m16_true_sparse}),
    with the matrices from \Cref{alg:mc_lbcs} (\ref{curve:lbcs_p1_m16_true_sparse}),
    and with the matrices from \Cref{alg:learning} (\ref{curve:learned_p1_m16_true_sparse}).
}
\label{fig:true_basis_16}
\end{figure}

The left plot in \Cref{fig:true_basis_16} visualizes evaluation results with model~\eqref{eq:model1} for $ m = 16 $.
Generally, we can see that recovering channels with sparsity $ p = 10 $ from $ m = 16 $ measurements is not possible:
the relative \ac{mse} is larger than one.
The case $ p = 5 $ shows better performance and the case $ p = 1 $ seems to show perfect recovery up to the remaining error which is due to noise.
Given the fact that we try to recover $ p $ nonzero entries of an $ n = 128 $-dimensional sparse vector from only $ m = 16 $ measurements,
the qualitative behavior of the displayed curves can be expected.

It is worth pointing out that while the performance of the three displayed methods of generating measurement matrices is aligned in the case $ p = 1 $,
we only have one matrix per \ac{snr} when \Cref{alg:mc_lbcs} is used
and we only have a single matrix for the whole curve when the proposed \Cref{alg:learning} is used.
In contrast, the random matrices approach generates a new random matrix for every channel.
All three methods lead to equally bad performance for $ p = 10 $ where the limiting factor seems to be the too small number of measurements $ m = 16 $.
Further, for $ p = 5 $, both \Cref{alg:mc_lbcs,alg:learning} improve on random matrices.

The right plot in \Cref{fig:true_basis_16} visualizes the same setting but with model~\eqref{eq:model2}.
According to compressive sensing theory, there should not be a performance difference
between the two channel models~\eqref{eq:model1} or~\eqref{eq:model2} when random matrices are employed.
This is due to the fact that random matrices enjoy a universality property
which means that they are incoherent to any basis~\cite{bookFoRa13}.
However, the case $ p = 1 $ in the right plot clearly behaves differently from the corresponding case in the left plot.
We can understand this behavior when we take into account that both the random constant modulus matrices $ \mbA $ as well as the \ac{dft} basis $ \mbF $ have
entries of the form $ e^{\imag \phi} $.
This can lead to a large coherence between $ \mbA $ and $ \mbF $ which can degrade the performance.
Remarkably, \Cref{alg:mc_lbcs,alg:learning} show equally good performance in both cases:
the respective training processes are able to find incoherent enough measurement matrices.
Even in the simple case $ p = 1 $, the observed behavior of the algorithms is already interesting.
It emphasizes the difference in the obtained matrices.

%
% true_sparse and basis_sparse, m = 32
%
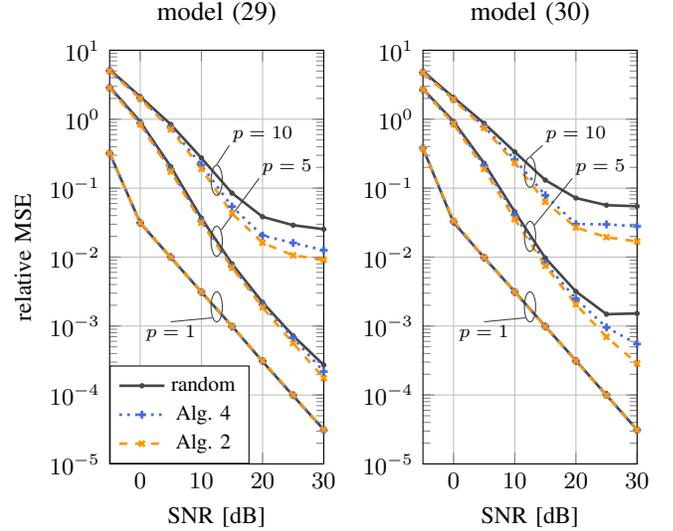
\begin{figure}[!t]
\centering
    \begin{tikzpicture}
        \begin{axis}[
            ymode=log,
            height=\plotheightsparse,
            width=\plotwidthsparse,
            ticks=both,
            xmin=-5, xmax=30,
            ymin=0.00001, ymax=10,
            xlabel={SNR [dB]},
            ylabel={relative MSE},
            title={model~\eqref{eq:model1}},
            xmajorgrids,
            ymajorgrids,
            legend style={at={(0.0, 0.0)}, anchor=south west, legend columns=1},
        ]

        % random
        % p = 1
        \addplot[omp1, discard if not={algorithm}{random}]
            table [ignore chars=",x=snr,y=mean,col sep=comma]
            {csv/compare_all/true_sparse/nAntennas_128_nPaths_1_nObservations_32.csv};
        \label{curve:random_p1_m32_true_sparse}
        \addlegendentry{\legendomp}

        % random
        % p = 5
        \addplot[forget plot, omp1, discard if not={algorithm}{random}]
            table [ignore chars=",x=snr,y=mean,col sep=comma]
            {csv/compare_all/true_sparse/nAntennas_128_nPaths_5_nObservations_32.csv};

        % random
        % p = 10
        \addplot[forget plot, omp1, discard if not={algorithm}{random}]
            table [ignore chars=",x=snr,y=mean,col sep=comma]
            {csv/compare_all/true_sparse/nAntennas_128_nPaths_10_nObservations_32.csv};

        % lbcs
        % p = 1
        \addplot[lbcs1, discard if not={algorithm}{lbcs}]
            table [ignore chars=",x=snr,y=mean,col sep=comma]
            {csv/compare_all/true_sparse/nAntennas_128_nPaths_1_nObservations_32.csv};
        \label{curve:lbcs_p1_m32_true_sparse}
        \addlegendentry{\legendlbcs}

        % lbcs
        % p = 5
        \addplot[forget plot, lbcs1, discard if not={algorithm}{lbcs}]
            table [ignore chars=",x=snr,y=mean,col sep=comma]
            {csv/compare_all/true_sparse/nAntennas_128_nPaths_5_nObservations_32.csv};

        % lbcs
        % p = 10
        \addplot[forget plot, lbcs1, discard if not={algorithm}{lbcs}]
            table [ignore chars=",x=snr,y=mean,col sep=comma]
            {csv/compare_all/true_sparse/nAntennas_128_nPaths_10_nObservations_32.csv};

        % learned
        % p = 1
        \addplot[learned1, discard if not={algorithm}{learned}]
            table [ignore chars=",x=snr,y=mean,col sep=comma]
            {csv/compare_all/true_sparse/nAntennas_128_nPaths_1_nObservations_32.csv};
        \label{curve:learned_p1_m32_true_sparse}
        \addlegendentry{\legendlearned}

        % learned
        % p = 5
        \addplot[forget plot, learned1, discard if not={algorithm}{learned}]
            table [ignore chars=",x=snr,y=mean,col sep=comma]
            {csv/compare_all/true_sparse/nAntennas_128_nPaths_5_nObservations_32.csv};

        % learned
        % p = 10
        \addplot[forget plot, learned1, discard if not={algorithm}{learned}]
            table [ignore chars=",x=snr,y=mean,col sep=comma]
            {csv/compare_all/true_sparse/nAntennas_128_nPaths_10_nObservations_32.csv};

        \draw (axis cs:12.5,0.13) node[ellipse, darkgray, inner xsep=0pt, inner ysep=0pt, minimum width=0.15cm, minimum height=0.4cm, draw] (e10) {};
        \node[inner sep=1pt] (n10) at (axis cs:20,0.6) {\scriptsize $ p = 10 $};
        \draw[darkgray] (n10.south west) -- (e10.north);

        \draw (axis cs:12.5,0.017) node[ellipse, darkgray, inner xsep=0pt, inner ysep=0pt, minimum width=0.15cm, minimum height=0.4cm, draw] (e5) {};
        \node[inner sep=1pt] (n5) at (axis cs:24,0.2) {\scriptsize $ p = 5 $};
        \draw[darkgray] (n5.south west) -- (e5.north east);

        \draw (axis cs:12.5,0.0019) node[ellipse, darkgray, inner xsep=0pt, inner ysep=0pt, minimum width=0.15cm, minimum height=0.4cm, draw] (e1) {};
        \node[inner sep=1pt] (n1) at (axis cs:5,0.0008) {\scriptsize $ p = 1 $};
        \draw[darkgray] (n1.north) -- (e1.south west);

        \end{axis}
    \end{tikzpicture}
    %%%%%%%%%%
    \begin{tikzpicture}
        \begin{axis}[
            ymode=log,
            height=\plotheightsparse,
            width=\plotwidthsparse,
            ticks=both,
            xmin=-5, xmax=30,
            ymin=0.00001, ymax=10,
            xlabel={SNR [dB]},
            title={model~\eqref{eq:model2}},
            xmajorgrids,
            ymajorgrids,
            legend style={at={(0.0, 0.0)}, anchor=south west, legend columns=1},
        ]

        % random
        % p = 1
        \addplot[omp1, discard if not={algorithm}{random}]
            table [ignore chars=",x=snr,y=mean,col sep=comma]
            {csv/compare_all/basis_sparse/nAntennas_128_nPaths_1_nObservations_32.csv};

        % random
        % p = 5
        \addplot[forget plot, omp1, discard if not={algorithm}{random}]
            table [ignore chars=",x=snr,y=mean,col sep=comma]
            {csv/compare_all/basis_sparse/nAntennas_128_nPaths_5_nObservations_32.csv};

        % random
        % p = 10
        \addplot[forget plot, omp1, discard if not={algorithm}{random}]
            table [ignore chars=",x=snr,y=mean,col sep=comma]
            {csv/compare_all/basis_sparse/nAntennas_128_nPaths_10_nObservations_32.csv};

        % lbcs
        % p = 1
        \addplot[lbcs1, discard if not={algorithm}{lbcs}]
            table [ignore chars=",x=snr,y=mean,col sep=comma]
            {csv/compare_all/basis_sparse/nAntennas_128_nPaths_1_nObservations_32.csv};

        % lbcs
        % p = 5
        \addplot[forget plot, lbcs1, discard if not={algorithm}{lbcs}]
            table [ignore chars=",x=snr,y=mean,col sep=comma]
            {csv/compare_all/basis_sparse/nAntennas_128_nPaths_5_nObservations_32.csv};

        % lbcs
        % p = 10
        \addplot[forget plot, lbcs1, discard if not={algorithm}{lbcs}]
            table [ignore chars=",x=snr,y=mean,col sep=comma]
            {csv/compare_all/basis_sparse/nAntennas_128_nPaths_10_nObservations_32.csv};

        % learned
        % p = 1
        \addplot[learned1, discard if not={algorithm}{learned}]
            table [ignore chars=",x=snr,y=mean,col sep=comma]
            {csv/compare_all/basis_sparse/nAntennas_128_nPaths_1_nObservations_32.csv};

        % learned
        % p = 5
        \addplot[forget plot, learned1, discard if not={algorithm}{learned}]
            table [ignore chars=",x=snr,y=mean,col sep=comma]
            {csv/compare_all/basis_sparse/nAntennas_128_nPaths_5_nObservations_32.csv};

        % learned
        % p = 10
        \addplot[forget plot, learned1, discard if not={algorithm}{learned}]
            table [ignore chars=",x=snr,y=mean,col sep=comma]
            {csv/compare_all/basis_sparse/nAntennas_128_nPaths_10_nObservations_32.csv};

        \draw (axis cs:12.5,0.17) node[ellipse, darkgray, inner xsep=0pt, inner ysep=0pt, minimum width=0.15cm, minimum height=0.4cm, draw] (e10) {};
        \node[inner sep=1pt] (n10) at (axis cs:20,0.6) {\scriptsize $ p = 10 $};
        \draw[darkgray] (n10.south west) -- (e10.north);

        \draw (axis cs:12.5,0.02) node[ellipse, darkgray, inner xsep=0pt, inner ysep=0pt, minimum width=0.15cm, minimum height=0.4cm, draw] (e5) {};
        \node[inner sep=1pt] (n5) at (axis cs:24,0.2) {\scriptsize $ p = 5 $};
        \draw[darkgray] (n5.south west) -- (e5.north east);

        \draw (axis cs:12.5,0.0019) node[ellipse, darkgray, inner xsep=0pt, inner ysep=0pt, minimum width=0.15cm, minimum height=0.4cm, draw] (e1) {};
        \node[inner sep=1pt] (n1) at (axis cs:5,0.0008) {\scriptsize $ p = 1 $};
        \draw[darkgray] (n1.north) -- (e1.south west);

        \end{axis}
    \end{tikzpicture}
\caption{%
    Evaluation with channel models~\eqref{eq:model1} (left) and~\eqref{eq:model2} (right).
    The constant modulus matrices $ \mbA $ have size $ 32 \times 128 $ and the sparsity $ p \in \{1, 5, 10\} $ is varied.
    Evaluation is shown with random matrices (\ref{curve:random_p1_m32_true_sparse}),
    with the matrices from \Cref{alg:mc_lbcs} (\ref{curve:lbcs_p1_m32_true_sparse}),
    and with the matrices from \Cref{alg:learning} (\ref{curve:learned_p1_m32_true_sparse}).
}
\label{fig:true_basis_32}
\end{figure}

\Cref{fig:true_basis_32} repeats the experiments for $ m = 32 $.
It can be seen that now all cases $ p \in \{1, 5, 10\} $ show a relative \ac{mse} below one
which can be interpreted to mean that the number of measurements $ m $ starts to be large enough.
Additionally, we observe that the case $ p = 5 $ starts to be parallel to $ p = 1 $
which is to be expected once $ m $ is large enough.
Further, the incoherence problem for model~\eqref{eq:model2} in contrast to model~\eqref{eq:model1} is not as pronounced anymore
which further supports the provided explanation.
Still, a degradation can be observed in the right plot for $ p = 5 $ and again \Cref{alg:mc_lbcs,alg:learning} can alleviate the problem.
Lastly, we see how the two learning-based methods improve on random matrices in particular for $ p = 10 $.

We also simulated $ m = 64 $ and $ m = 96 $ where all curves essentially are straight lines and all methods are aligned (not shown here).
At this point, we would like to emphasize again that this is a satisfying result
since, e.g., the proposed \Cref{alg:learning} gives us a single measurement matrix that we use for all \acp{snr},
which may offer attractive options for potential technical implementations.

%%%%%%%%%%%%%%%%%%%%%%%%%%%%%%%%%%%%%%%%%%%%%%%%%%%%%%%%%%%%%%%%%%%%%%%%%%%%%%%
\subsection{Discussion of Results for Approximately Sparse Vectors}\label{sec:discussion_approx_sparse}

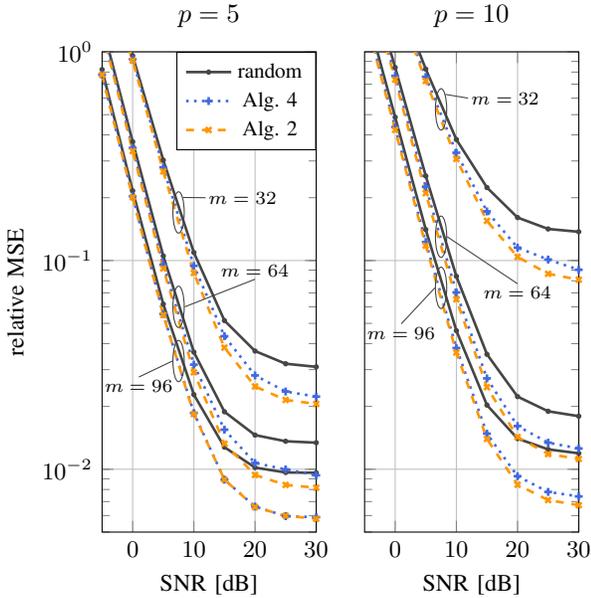
\begin{figure}[!t]
\centering
    \begin{tikzpicture}
        \begin{axis}[
            ymode=log,
            height=\plotheightdictionary,
            width=\plotwidthdictionary,
            ticks=both,
            xmin=-5, xmax=30,
            ymin=0.005, ymax=1,
            xlabel={SNR [dB]},
            ylabel={relative MSE},
            title={$ p = 5 $},
            xmajorgrids,
            ymajorgrids,
            legend style={at={(1.0, 1.0)}, anchor=north east, legend columns=1},
        ]

        % random
        % n_observations = 32
        \addplot[omp1, discard if not={algorithm}{random}]
            table [ignore chars=",x=snr,y=mean,col sep=comma]
            {csv/compare_all/dictionary_sparse/nAntennas_128_nPaths_5_nObservations_32.csv};
        \label{curve:random_p5_m32}
        \addlegendentry{\legendomp}

        % random
        % n_observations = 64
        \addplot[forget plot, omp1, discard if not={algorithm}{random}]
            table [ignore chars=",x=snr,y=mean,col sep=comma]
            {csv/compare_all/dictionary_sparse/nAntennas_128_nPaths_5_nObservations_64.csv};

        % random
        % n_observations = 96
        \addplot[forget plot, omp1, discard if not={algorithm}{random}]
            table [ignore chars=",x=snr,y=mean,col sep=comma]
            {csv/compare_all/dictionary_sparse/nAntennas_128_nPaths_5_nObservations_96.csv};

        % lbcs
        % n_observations = 32
        \addplot[lbcs1, discard if not={algorithm}{lbcs}]
            table [ignore chars=",x=snr,y=mean,col sep=comma]
            {csv/compare_all/dictionary_sparse/nAntennas_128_nPaths_5_nObservations_32.csv};
        \label{curve:lbcs_p5_m32}
        \addlegendentry{\legendlbcs}

        % lbcs
        % n_observations = 64
        \addplot[forget plot, lbcs1, discard if not={algorithm}{lbcs}]
            table [ignore chars=",x=snr,y=mean,col sep=comma]
            {csv/compare_all/dictionary_sparse/nAntennas_128_nPaths_5_nObservations_64.csv};

        % lbcs
        % n_observations = 96
        \addplot[forget plot, lbcs1, discard if not={algorithm}{lbcs}]
            table [ignore chars=",x=snr,y=mean,col sep=comma]
            {csv/compare_all/dictionary_sparse/nAntennas_128_nPaths_5_nObservations_96.csv};

        % learned
        % n_observations = 32
        \addplot[learned1, discard if not={algorithm}{learned}]
            table [ignore chars=",x=snr,y=mean,col sep=comma]
            {csv/compare_all/dictionary_sparse/nAntennas_128_nPaths_5_nObservations_32.csv};
        \label{curve:learned_p5_m32}
        \addlegendentry{\legendlearned}

        % learned
        % n_observations = 64
        \addplot[forget plot, learned1, discard if not={algorithm}{learned}]
            table [ignore chars=",x=snr,y=mean,col sep=comma]
            {csv/compare_all/dictionary_sparse/nAntennas_128_nPaths_5_nObservations_64.csv};

        % learned
        % n_observations = 96
        \addplot[forget plot, learned1, discard if not={algorithm}{learned}]
            table [ignore chars=",x=snr,y=mean,col sep=comma]
            {csv/compare_all/dictionary_sparse/nAntennas_128_nPaths_5_nObservations_96.csv};

        \draw (axis cs:7.5,0.17) node[ellipse, darkgray, inner xsep=0pt, inner ysep=0pt, minimum width=0.15cm, minimum height=0.55cm, draw] (e32) {};
        \node[inner sep=1pt] (n32) at (axis cs:18,0.2) {\scriptsize $ m = 32 $};
        \draw[darkgray] (n32.west) -- (e32.north east);

        \draw (axis cs:7.5,0.06) node[ellipse, darkgray, inner xsep=0pt, inner ysep=0pt, minimum width=0.15cm, minimum height=0.55cm, draw] (e64) {};
        \node[inner sep=1pt] (n64) at (axis cs:20,0.09) {\scriptsize $ m = 64 $};
        \draw[darkgray] (n64.south) -- (e64.east);

        \draw (axis cs:7.5,0.033) node[ellipse, darkgray, inner xsep=0pt, inner ysep=0pt, minimum width=0.15cm, minimum height=0.55cm, draw] (e96) {};
        \node[inner sep=1pt] (n96) at (axis cs:1,0.025) {\scriptsize $ m = 96 $};
        \draw[darkgray] (n96.north) -- (e96.west);

        \end{axis}
    \end{tikzpicture}
    %%%%%%%%%%
    \begin{tikzpicture}
        \begin{axis}[
            ymode=log,
            height=\plotheightdictionary,
            width=\plotwidthdictionary,
            ticks=both,
            xmin=-5, xmax=30,
            ymin=0.005, ymax=1,
            xlabel={SNR [dB]},
            title={$ p = 10 $},
            xmajorgrids,
            ymajorgrids,
            yticklabels={,,},
            legend style={at={(1.0, 1.0)}, anchor=north east, legend columns=1},
        ]

        % random
        % n_observations = 32
        \addplot[omp1, discard if not={algorithm}{random}]
            table [ignore chars=",x=snr,y=mean,col sep=comma]
            {csv/compare_all/dictionary_sparse/nAntennas_128_nPaths_10_nObservations_32.csv};

        % random
        % n_observations = 64
        \addplot[forget plot, omp1, discard if not={algorithm}{random}]
            table [ignore chars=",x=snr,y=mean,col sep=comma]
            {csv/compare_all/dictionary_sparse/nAntennas_128_nPaths_10_nObservations_64.csv};

        % random
        % n_observations = 96
        \addplot[forget plot, omp1, discard if not={algorithm}{random}]
            table [ignore chars=",x=snr,y=mean,col sep=comma]
            {csv/compare_all/dictionary_sparse/nAntennas_128_nPaths_10_nObservations_96.csv};

        % lbcs
        % n_observations = 32
        \addplot[lbcs1, discard if not={algorithm}{lbcs}]
            table [ignore chars=",x=snr,y=mean,col sep=comma]
            {csv/compare_all/dictionary_sparse/nAntennas_128_nPaths_10_nObservations_32.csv};

        % lbcs
        % n_observations = 64
        \addplot[forget plot, lbcs1, discard if not={algorithm}{lbcs}]
            table [ignore chars=",x=snr,y=mean,col sep=comma]
            {csv/compare_all/dictionary_sparse/nAntennas_128_nPaths_10_nObservations_64.csv};

        % lbcs
        % n_observations = 96
        \addplot[forget plot, lbcs1, discard if not={algorithm}{lbcs}]
            table [ignore chars=",x=snr,y=mean,col sep=comma]
            {csv/compare_all/dictionary_sparse/nAntennas_128_nPaths_10_nObservations_96.csv};

        % learned
        % n_observations = 32
        \addplot[learned1, discard if not={algorithm}{learned}]
            table [ignore chars=",x=snr,y=mean,col sep=comma]
            {csv/compare_all/dictionary_sparse/nAntennas_128_nPaths_10_nObservations_32.csv};

        % learned
        % n_observations = 64
        \addplot[forget plot, learned1, discard if not={algorithm}{learned}]
            table [ignore chars=",x=snr,y=mean,col sep=comma]
            {csv/compare_all/dictionary_sparse/nAntennas_128_nPaths_10_nObservations_64.csv};

        % learned
        % n_observations = 96
        \addplot[forget plot, learned1, discard if not={algorithm}{learned}]
            table [ignore chars=",x=snr,y=mean,col sep=comma]
            {csv/compare_all/dictionary_sparse/nAntennas_128_nPaths_10_nObservations_96.csv};

        \draw (axis cs:7.5,0.53) node[ellipse, darkgray, inner xsep=0pt, inner ysep=0pt, minimum width=0.15cm, minimum height=0.55cm, draw] (e32) {};
        \node[inner sep=1pt] (n32) at (axis cs:18,0.6) {\scriptsize $ m = 32 $};
        \draw[darkgray] (n32.west) -- (e32.north east);

        \draw (axis cs:7.5,0.13) node[ellipse, darkgray, inner xsep=0pt, inner ysep=0pt, minimum width=0.15cm, minimum height=0.55cm, draw] (e64) {};
        \node[inner sep=1pt] (n64) at (axis cs:20,0.07) {\scriptsize $ m = 64 $};
        \draw[darkgray] (n64.north) -- (e64.east);

        \draw (axis cs:7.5,0.074) node[ellipse, darkgray, inner xsep=0pt, inner ysep=0pt, minimum width=0.15cm, minimum height=0.55cm, draw] (e96) {};
        \node[inner sep=1pt] (n96) at (axis cs:1,0.045) {\scriptsize $ m = 96 $};
        \draw[darkgray] (n96.north) -- (e96.west);

        \end{axis}
    \end{tikzpicture}
\caption{%
    Evaluation with channel model~\eqref{eq:model3}.
    The constant modulus matrices $ \mbA $ have size $ m \times 128 $ where $ m \in \{32, 64, 96\} $.
    The number of paths is $ p = 5 $ (left) and $ p = 10 $ (right).
    Evaluation is shown with random matrices (\ref{curve:random_p5_m32}),
    with the matrices from \Cref{alg:mc_lbcs} (\ref{curve:lbcs_p5_m32}) and from \Cref{alg:learning} (\ref{curve:learned_p5_m32}).
}
\label{fig:dictionary_5_10}
\end{figure}

This subsection concentrates on channel model~\eqref{eq:model3}.
\Cref{fig:dictionary_5_10} shows the evaluation of matrices $ \mbA \in \C^{m\times 128} $ with $ m \in \{32, 64, 96\} $
for $ p = 5 $ (left) and $ p = 10 $ (right) paths.
We see that all curves are eventually constant with increasing \ac{snr}.
The reason for this phenomenon is that the channels are only approximately sparse: $ \mbh \approx \mbPsi_L \mbs $.
Thus, even in the noiseless case, perfect recovery is not possible
whenever angles $ \theta_k $ of some steering vectors lie in between two grid points of the dictionary $ \mbPsi_L $.
There exist different approaches to combat this problem.
On overview of grid-less methods can, e.g., be found in~\cite[Chapter~11]{YaLiStXi18}.
However, since we are only interested in studying the behavior of different measurement matrices
rather than in the actual channel estimation itself, it suffices for our purposes to work with the on-grid \ac{omp} algorithm.

In all cases, \Cref{alg:mc_lbcs,alg:learning} can significantly improve on random matrices.
Consider for example the cases $ m \in \{64, 96\} $ in the left plot of \Cref{fig:dictionary_5_10}.
The performance of the matrices for $ m = 64 $ of \Cref{alg:mc_lbcs,alg:learning} reaches the performance of random matrices for $ m = 96 $.
In this sense, the learning processes lead to an improvement of the matrix $ \mbA $ which corresponds to a gain of 32 measurements.
Additionally, it can be kept in mind that \Cref{alg:mc_lbcs,alg:learning} only need one matrix
(per \ac{snr} in case of \Cref{alg:mc_lbcs}) in contrast to the random matrices approach.

%
% dictionary_sparse, p = 5 and p = 10
%

We see a considerable performance degradation for the case $ m = 32 $
when $ p $ is increased from 5 to 10 (compare left and right plots in \Cref{fig:dictionary_5_10}).
This may again be interpreted as having a too small number of measurements $ m $.

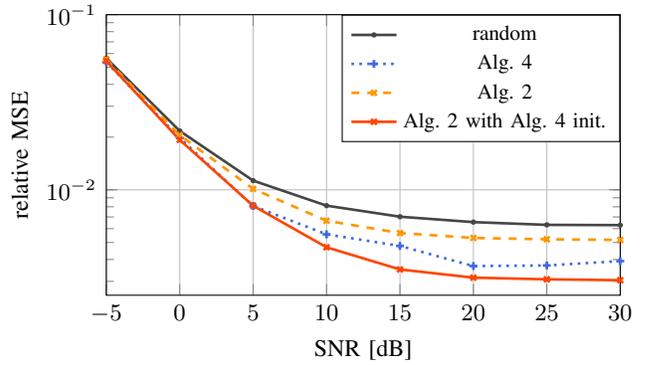
\begin{figure}[!t]
\centering
    \begin{tikzpicture}
        \begin{axis}[
            ymode=log,
            height=\plotheight,
            width=\plotwidth,
            ticks=both,
            xmin=-5, xmax=30,
            ymin=0.0025, ymax=0.1,
            xlabel={SNR [dB]},
            ylabel={relative MSE},
            xmajorgrids,
            ymajorgrids,
            legend style={at={(1.0, 1.0)}, anchor=north east, legend columns=1},
        ]

        % random
        % n_observations = 96
        \addplot[omp1, discard if not={algorithm}{random}]
            table [ignore chars=",x=snr,y=mean,col sep=comma]
            {csv/compare_all/dictionary_sparse/nAntennas_128_nPaths_1_nObservations_96.csv};
        \label{curve:random_p1_m96}
        \addlegendentry{\legendomp}

        % lbcs
        % n_observations = 96
        \addplot[lbcs1, discard if not={algorithm}{lbcs}]
            table [ignore chars=",x=snr,y=mean,col sep=comma]
            {csv/compare_all/dictionary_sparse/nAntennas_128_nPaths_1_nObservations_96.csv};
        \label{curve:lbcs_p1_m96}
        \addlegendentry{\legendlbcs}

        % learned
        % n_observations = 96
        \addplot[learned1, discard if not={algorithm}{learned}]
            table [ignore chars=",x=snr,y=mean,col sep=comma]
            {csv/compare_all/dictionary_sparse/nAntennas_128_nPaths_1_nObservations_96.csv};
        \label{curve:learned_p1_m96}
        \addlegendentry{\legendlearned}

        % learned, lbcs init
        % n_observations = 96
        \addplot[learned3]
            table [ignore chars=",x=snr,y=mean,col sep=comma]
            {csv/lbcs_init/nAntennas_128_nPaths_1_nObservations_96.csv};
        \label{curve:init}
        \addlegendentry{\legendinit}

        \end{axis}
    \end{tikzpicture}
\caption{%
    Evaluation with channel model~\eqref{eq:model3}.
    The constant modulus matrices $ \mbA $ have size $ 96 \times 128 $.
    The number of paths is $ p = 1 $.
    Evaluation is shown with random matrices (\ref{curve:random_p1_m96}), and
    with the matrices from \Cref{alg:mc_lbcs} (\ref{curve:lbcs_p1_m96}) and from \Cref{alg:learning} (\ref{curve:learned_p1_m96}).
    Further, the result of initializing \Cref{alg:learning} with \Cref{alg:mc_lbcs} is depicted (\ref{curve:init}).
}
\label{fig:lbcs_init}
\end{figure}

So far, in all considered cases, \Cref{alg:mc_lbcs,alg:learning} were at least as good or better than random matrices
and the proposed \Cref{alg:learning} was at least as good or better than \Cref{alg:mc_lbcs}.
However, this changes if we consider $ m = 96 $ and $ p = 1 $, which is displayed in \Cref{fig:lbcs_init}.
In theory, every matrix which can be found via \Cref{alg:mc_lbcs} is a matrix which could also be found via \Cref{alg:learning},
so that we could expect the performance of \Cref{alg:learning}
to be comparable to the one of \Cref{alg:mc_lbcs}.
In addition to the discussed hyperparameters, which we determine via random search,
another important factor of \Cref{alg:learning} is the initialization of the matrix $ \mbPhi $.
In all considered numerical results, for initialization, every element was drawn independently and uniformly in the interval $ [0, 2\pi] $.
It is now an interesting experiment to initialize \Cref{alg:learning} with \Cref{alg:mc_lbcs} and to observe
whether the learning procedure does at least not degrade the quality of the final matrix.

This was done in \Cref{fig:lbcs_init} where we used \Cref{alg:mc_lbcs} with
$ \mr{SNR} = 20 $ dB to initialize \Cref{alg:learning} before
we run the 64 random searches as described in \Cref{sec:implementation_details}.
The corresponding curve reveals that the two algorithms harmonize well.
Finding a good initialization is a general issue of gradient descent algorithms.
The observation in \Cref{fig:lbcs_init}
 suggests to make initialization via \Cref{alg:mc_lbcs} part of the random search:
in one half of all cases, the initialization would be random, in the other half, the initialization would be based on \Cref{alg:mc_lbcs}.

%
% dictionary_sparse
%

%%%%%%%%%%%%%%%%%%%%%%%%%%%%%%%%%%%%%%%%%%%%%%%%%%%%%%%%%%%%%%%%%%%%%%%%%%%%%%%
\subsection{Training Data Normalization}\label{sec:data_normalization}

Lastly, we want to discuss training data preprocessing for \Cref{alg:learning}.
The initial idea described in \Cref{sec:main} is to regard $ \mbA \in \C^{m\times n} $ as a mapping
from one hypersphere (in $ \C^n $) to another hypersphere (in $ \C^m $).
This implies to normalize the training data $ \{ \mbh_i \}_{i=1}^{\Ttrain} $ via
\begin{equation}\label{eq:normalization}
    \htilde_i = \frac{\mbh_i}{\|\mbh_i\|}, \quad \forall i = 1, 2, \dots, \Ttrain
\end{equation}
such that $ \htilde_i $ represents a point on the hypersphere in $ \C^n $.

However, if the distribution of $ \mbA \htilde $ now equaled the distribution of $ \mbu $
(cf.~\eqref{eq:abstract_problem}), we would have $ \| \mbA \htilde \| = 1 $
which means we would have found a matrix with \ac{rip} constant $ \delta $ equal to zero,
see~\eqref{eq:rip_divided}.
This can be expected to be difficult if not impossible.
In particular, if the number of rows, $ m $, is significantly smaller than the number of columns, $ n $,
then $ \delta $ can be expected to be relatively large.
On the other hand, if $ m $ is larger,
the matrix $ \mbA $ is closer to a square matrix and it is easier to find an (approximate) isometry.
With this intuition in mind, the optimization problem may be difficult to solve
if the data is normalized such that every $ \htilde_i $ lies exactly on the hypersphere in $ \C^n $,
i.e., if we have $ \| \htilde_i \| = 1 $.
To alleviate the problem, we heuristically introduce some variation in the norm of the training data.
Specifically, we normalize the training data as described in equation~\eqref{eq:normalization_expec},
where the denominator approximates $ \sqrt{\expec[\|\mbh\|^2]} $.
Consequently, the training data points are only ``on average'' on the hypersphere in $ \C^n $.
Now, the distribution $ \mbA \htilde $ can be equal to the distribution of $ \mbu $,
but we would not have $ \delta = 0 $.

\begin{figure}[!t]
\centering
    \begin{tikzpicture}
        \begin{axis}[
            ymode=log,
            height=\plotheightdelta,
            width=\plotwidthdelta,
            ticks=both,
            xmin=-5, xmax=30,
            ymin=0.005, ymax=2,
            xlabel={SNR [dB]},
            ylabel={relative MSE},
            xmajorgrids,
            ymajorgrids,
            legend style={at={(1.0, 1.0)}, anchor=north east, legend columns=1, cells={anchor=west}},
        ]

        % omp
        % n_observations = 16
        \addplot[style1]
            table [ignore chars=",x=snr,y=mean,col sep=comma]
            {csv/with_delta/nAntennas_128_nPaths_1_nObservations_16_omp.csv};
        \label{curve:m16random}
        \addlegendentry{$ m = 16 $, \legendomp}

        % n_observations = 16
        % delta = 0.0
        \addplot[style2, discard if not={delta}{0.0}]
            table [ignore chars=",x=snr,y=mean,col sep=comma]
            {csv/with_delta/nAntennas_128_nPaths_1_nObservations_16_mean.csv};
        \label{curve:m16delta0}
        \addlegendentry{$ m = 16, \| \cdot \| $}

        % n_observations = 16
        % delta = 1.0
        \addplot[style3, discard if not={delta}{1.0}]
            table [ignore chars=",x=snr,y=mean,col sep=comma]
            {csv/with_delta/nAntennas_128_nPaths_1_nObservations_16_mean.csv};
        \label{curve:m16delta1}
        \addlegendentry{$ m = 16, \sqrt{\expec[\| \cdot \|^2]} $}

        % omp
        % n_observations = 96
        \addplot[style4]
            table [ignore chars=",x=snr,y=mean,col sep=comma]
            {csv/with_delta/nAntennas_128_nPaths_10_nObservations_96_omp.csv};
        \label{curve:m96random}
        \addlegendentry{$ m = 96 $, \legendomp}
        
        % n_observations = 96
        % delta = 0.0
        \addplot[style5, discard if not={delta}{0.0}]
            table [ignore chars=",x=snr,y=mean,col sep=comma]
            {csv/with_delta/nAntennas_128_nPaths_10_nObservations_96_mean.csv};
        \label{curve:m96delta0}
        \addlegendentry{$ m = 96, \| \cdot \| $}

        % n_observations = 96
        % delta = 1.0
        \addplot[style6, discard if not={delta}{1.0}]
            table [ignore chars=",x=snr,y=mean,col sep=comma]
            {csv/with_delta/nAntennas_128_nPaths_10_nObservations_96_mean.csv};
        \label{curve:m96delta1}
        \addlegendentry{$ m = 96, \sqrt{\expec[\| \cdot \|^2]} $}

        \end{axis}
    \end{tikzpicture}
\caption{%
    Dashed curves show evaluation with model~\eqref{eq:model3} for $ m = 16 $ and $ p = 1 $, solid curves for $ m = 96 $ and $ p = 10 $.
    Evaluation with random constant modulus matrices is displayed in dark gray (\ref{curve:m16random}, \ref{curve:m96random}).
    Blue curves (\ref{curve:m16delta0}, \ref{curve:m96delta0}) refer to learning a matrix with training data normalization $ \mbh/\|\mbh\| $,
    orange curves (\ref{curve:m16delta1}, \ref{curve:m96delta1}) refer to normalization $ \mbh/\sqrt{\expec[\|\mbh\|^2]} $.
}
\label{fig:different_deltas}
\end{figure}
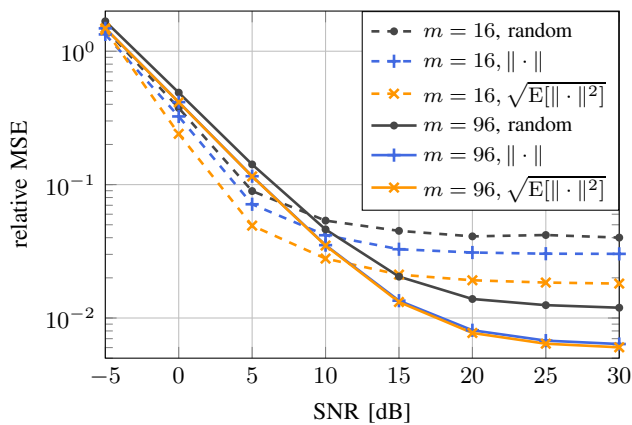

The effect of the different normalizations is studied in \Cref{fig:different_deltas}
where we consider two extreme cases: few measurements $ m = 16 $ but only a single path $ p = 1 $
and many measurements $ m = 96 $ with many paths $ p = 10 $.
It can be seen that the normalization~\eqref{eq:normalization_expec} leads to better results than~\eqref{eq:normalization}.
The difference is particularly pronounced in the case $ m = 16 $.
This fits the intuition provided above: as $ m $ decreases, $ \delta $ potentially needs to increase.
Hence, training with~\eqref{eq:normalization} is more restrictive in that case.
Motivated by \Cref{fig:different_deltas},
normalization~\eqref{eq:normalization_expec} was used in all other numerical experiments presented in this section.

%
% delta
%

%%%%%%%%%%%%%%%%%%%%%%%%%%%%%%%%%%%%%%%%%%%%%%%%%%%%%%%%%%%%%%%%%%%%%%%%%%%%%%%
%%%%%%%%%%%%%%%%%%%%%%%%%%%%%%%%%%%%%%%%%%%%%%%%%%%%%%%%%%%%%%%%%%%%%%%%%%%%%%%
\section{Conclusion}

The anchor point of our studies was the interpretation that a measurement matrix
which has the \ac{rip} constitutes (approximately) a mapping between hyperspheres.
In order to take measurement noise into account,
we proposed to determine the measurement matrix such
that it maps points uniformly onto the hypersphere in its range.
One of the \ac{mmd} metrics was used to measure the distance between a uniform distribution on the hypersphere
and the distribution in the range of the measurement matrix.
The recent success of \ac{mmd}-based machine learning applications motivated a solution of the optimization problem via stochastic gradient descent.
Further, it was shown that a modified method from the literature also leads to good performance
and that it can harmonize with our approach by providing a suitable initialization for the learning algorithm.

The algorithm's learning phase requires training data.
This training data may, for example, be obtained by means of a measurement campaign.
The learned matrix will then depend on the measured scenario.
In our experiments, this is reflected by the matrix's dependency on the number of propagation paths.
In this context, it might be interesting to investigate how the performance behaves when the scenario changes.
One might then attempt to employ a form of domain adaptation~\cite{bookGoBeCo16} to adaptively update the measurement matrix.

A notable property of the proposed learning algorithm is that it yields one measurement matrix
which can be employed for all \acp{snr}.
Numerical experiments demonstrate this property and the great performance of the proposed method.

In this paper, we focused on measurement matrices with a constant modulus constraint.
We translated the constraint into the system of equations shown in~\eqref{eq:stack_Ah}.
For an arbitrary (not necessarily constant modulus) matrix $ \mbC \in \C^{m\times n} $, \eqref{eq:stack_Ah} takes the form
\begin{equation}
    \begin{bmatrix}
        \Re(\mbC \htilde) \\
        \Im(\mbC \htilde)
    \end{bmatrix}
    =
    \begin{bmatrix}
        \Re(\mbC) & -\Im(\mbC) \\
        \Im(\mbC) & \Re(\mbC) \\
    \end{bmatrix}
    \begin{bmatrix}
        \Re(\htilde) \\
        \Im(\htilde)
    \end{bmatrix}.
\end{equation}
It can be seen that more general structural constraints can be incorporated into the design of $ \mbC $ as well.
For example, the authors of \cite{HaBaRaNo10} prove that random Toeplitz matrices can satisfy a \ac{rip} condition
and use this result to motivate the application of such matrices in a channel estimation context.
A Toeplitz measurement matrix can also be learned with the approach described in \Cref{sec:learning}.
To this end, weight tying can be employed~\cite{bookGoBeCo16}.
That is, parameters of $ \mbC $ would be shared between rows to enforce a Toeplitz structure.
Here it makes sense to normalize the rows during training.
Similarly, other structured matrices can be learned too.

%%%%%%%%%%%%%%%%%%%%%%%%%%%%%%%%%%%%%%%%%%%%%%%%%%%%%%%%%%%%%%%%%%%%%%%%%%%%%%%
%%%%%%%%%%%%%%%%%%%%%%%%%%%%%%%%%%%%%%%%%%%%%%%%%%%%%%%%%%%%%%%%%%%%%%%%%%%%%%%
\bibliographystyle{IEEEtran}
\bibliography{IEEEabrv,references}

\end{document}